\documentclass[11pt]{article}
\usepackage{jheppubmodnew}
\pdfoutput=1
\usepackage{hyperref}
\usepackage{psfrag}
\usepackage{array}
\usepackage{MnSymbol}
\usepackage{amsmath}
\usepackage{amsthm}
\usepackage{graphicx}
\usepackage{etoolbox}
	
\usepackage[punctsep]{collref}
\collectsep[]{;}
\usepackage{bm}
\usepackage{epsfig}
\usepackage{wrapfig}
\usepackage{tikz}
\usepackage{mathtools}
\usepackage{tensor}
\usepackage{cleveref}
\usepackage[makeroom]{cancel}
\usetikzlibrary{arrows,plotmarks,positioning,calc}

\newtheorem*{result}{Result}

\def\pb[#1,#2]{\{#1, #2\}}
\def\deb[#1,#2]{[#1,#2]_{\text{D.B.}}}

\def\l{\lambda}
\def\la{\lambda}

\def\a{\alpha}

\def\Or[#1]{\text{O}\!\left(#1\right)}
\def\dotl[#1,#2]{\left\langle #1,\, #2 \right\rangle}
\def\dotlb[#1,#2]{\left\langle #1,\, #2 \right\rangle}
\def\dotlm[#1,#2]{\left[ #1,\, #2 \right]}
\def\dotp[#1,#2]{(\vect{#1} \cdot\vect{#2})}
\def\aff[#1,#2]{\hat{#1}(#2)}

\def\n4sym{{\cal N}=4 SYM}
\def\>{\rangle}
\def\<{\langle}

\def\projsho[#1]{{\cal P}^{\text{sho}}_{#1}}
\def\transsho[#1,#2]{{\cal T}^{\text{sho}}_{#1,#2}}
\def\weight[#1,#2,#3]{\{(#1),#2,#3\}}
\def\ads[#1]{$\text{AdS}_{#1}$}

\newcommand{\be}{\begin{equation}}
\newcommand{\ee}{\end{equation}}
\newcommand{\ba}{\begin{align}}
\newcommand{\ea}{\end{align}}
\newcommand{\bs}{\begin{split}}
\def\sess\end{split}
\newcommand{\vect}[1]{{\boldsymbol{#1}}}

\newcommand{\bea}{\begin{eqnarray}}
\newcommand{\eea}{\end{eqnarray}}

\def \bes {\begin{equation*}}
\def \ees {\end{equation*}}

\def \b  {\beta}

\def \scrip{{\cal I}^{+}}

\def \scrippast{{\cal I}^{+}_{-}}

\def\alcut[#1]{{\cal A}_{#1, \epsilon}}
\def\alseg[#1,#2]{{\cal B}_{#1, #2}}

\def\supcharge[#1]{\{#1\}}

\def\projsupeig[#1]{{\cal P}_{{\ell, m}}[{#1}]}
\def\transop[#1, #2]{T_{\{#1\}, \{#2\}}}
\def\supket[#1]{|\{#1\} \rangle}
\def\supbra[#1]{\langle \{#1\} | }

\def\rsop[#1]{X_{#1}}

\def\projlow[#1,#2]{P_{{#1}<{#2}}}
\def\rd[#1]{^{(#1)}R}

\def\infz{\infty}
\def\confgrp{{\mathrm{SO}}(1,d+1)}
\def\sodminone{{\mathrm{SO}}(d-1)}
\def\reg{{\cal R}}

\def\coeff[#1,#2]{{\mathcal G}_{#1,#2}}
\def\dcoeff[#1,#2]{\delta {\mathcal G}_{#1,#2}}
\def\tcoeff[#1,#2]{{\widetilde{\mathcal G}}_{#1,#2}}
\def\lcoeff[#1,#2]{{\mathcal G}^{\lambda}_{#1,#2}}

\def\vdw{\text{vol(diff$\times$Weyl)}}
\def\cout{{\cal N}_1}
\def\cfp{{\cal N}_2}
\def\cfpp{{\cal N}_3}
\def\cprod{{\cal C}}
\def\ghosts{\text{gh}}
\def\resv{v}
\def\transv{\z}
\def\zmv{\zeta}
\newcommand\diffWeyl{diff $\times$ Weyl}
\def\vI{\vec{i}}
\def\vJ{\vec{j}}
\def\vK{\vec{k}}
\def\vL{\vec{\ell}}
\def\inv{\tilde}
\newcommand{\subf}[2]{%
	{\small\begin{tabular}[t]{@{}c@{}}
			#1\\#2
		\end{tabular}}%
	}

\def\bx{\inv{x}}
\def\bh{\inv{h}}
\def\bd{{\inv{\partial}}}

\def\pg{\paragraph}

\def\ie{\emph{i.e.} }

\def\nt{\notag}

\def\wt{\widetilde}

\def\R{\mathbb{R}}

\def\cD{\mathcal{D}}

\def\cL{\mathcal{L}}

\def\cN{\mathcal{N}}

\def\cR{\mathcal{R}}
\def\cS{\mathcal{S}}

\def\bD{\mathbb{D}}

\def\p{\partial}

\def\/{\over}

\def\rn{\rangle}
\def\ln{\langle}

\def\vphi{\varphi}
\def\a{\alpha}
\def\b{\beta}
\def\d{\delta}
\def\k{\kappa}
\def\g {\gamma}
\def\la {\lambda}

\def\z{\zeta}

\def\l{\ell}

\def\n {\nabla}

\def\D{\Delta}

\def\Om {\Omega}

\def\r{\mathrm}

\def\-{\\\notag}
\def\={&=&}

\newcommand{\bpm}{\begin{pmatrix}}
\newcommand{\epm}{\end{pmatrix}}

\newcommand{\bit}{\begin{itemize}}
\newcommand{\eit}{\end{itemize}}

\newcommand{\ben}{\begin{enumerate}}
\newcommand{\een}{\end{enumerate}}

\newcommand\bsp{\begin{split}}
\newcommand\esp{\end{split}}

\def\le{\left}
\def\ri{\right}

\def\l{\ell}

\def\qq{\qquad}

\newcommand{\enc}[1]{\left( #1\right)}

\newcommand{\encbr}[1]{\left\{#1\right\}}

\newcommand{\tD}{\text{D}}
\newcommand{\tW}{\text{W}}
\newcommand{\tB}{\text{B}}
\newcommand{\hdelta}{\hat{\delta}}

\def\langleQ{\llangle}
\def\rangleQ{\rrangle}
\def\qstate{\psi}

\newcommand\CC[2]{\llangle #1|#2|#1 \rrangle_{\text{CC}}}

\usepackage{comment}
\def\innerp[#1,#2]{\left({#1, #2} \right)}
\usepackage{cite}

\def\bD{{\bar{\D}}}

\newcommand{\changelocaltocdepth}[1]{%
  \addtocontents{toc}{\protect\setcounter{tocdepth}{#1}}%
  \setcounter{tocdepth}{#1}%
}
\title{Holography of information in de Sitter space}
\author[a]{Tuneer Chakraborty,}
\author[a,b]{Joydeep Chakravarty,}
\author[a]{Victor Godet,}
\author[a]{Priyadarshi Paul}
\author[a]{and Suvrat Raju}

\affiliation[a]{International Centre for Theoretical Sciences (ICTS-TIFR),\\ Tata Institute of Fundamental Research,
Shivakote, Hesaraghatta, Bengaluru 560089, India}

\affiliation[b]{Department of Physics, McGill University,\\
3600 Rue University, Montreal, H3A 2T8, QC Canada}

\emailAdd{tuneer.chakraborty@icts.res.in}
\emailAdd{joydeep.chakravarty@mail.mcgill.ca}
\emailAdd{victor.godet@icts.res.in}
\emailAdd{priyadarshi.paul@icts.res.in}
\emailAdd{suvrat@icts.res.in}
\date{}

\abstract{We study the natural norm on the space of solutions to the Wheeler-DeWitt equation in an asymptotically de Sitter spacetime. We propose that the norm is obtained by integrating the squared wavefunctional over field configurations and dividing by the volume of the diff-and-Weyl group.
We impose appropriate gauge conditions to fix the diff-and-Weyl redundancy and obtain a finite expression  for the norm using the Faddeev-Popov procedure. 
This leads to a ghost action that has zero modes corresponding to a residual conformal subgroup of the diff-and-Weyl group. By keeping track of these zero modes, we show that Higuchi's norm for group-averaged states emerges from our prescription in the nongravitational limit. We apply our formalism to  cosmological correlators and propose that they should be understood as gauge-fixed observables. We identify the symmetries of these observables.  In a nongravitational theory, it is necessary to specify such correlators everywhere on a Cauchy slice to identify a state in the Hilbert space. In a theory of quantum gravity, we demonstrate a version of the principle of holography of information:  cosmological correlators in an arbitrarily small region suffice to completely specify the state.}

\begin{document}
\maketitle
%*********** Introduction *******

\section{Introduction}
It is known that both in AdS and in flat space, quantum gravity localizes information very differently from nongravitational quantum field theories and manifests the principle of holography of information \cite{Laddha:2020kvp,Chowdhury:2020hse,Raju:2020smc,Chowdhury:2021nxw,Raju:2021lwh,Chakravarty:2023cll,deMelloKoch:2022sul}. In AdS, all information on a Cauchy slice is available near its boundary, as is well known from AdS/CFT but can also be shown directly from the gravitational theory. In flat space, it was shown in \cite{Laddha:2020kvp} that all information that can be obtained on future null infinity can also be obtained on its past boundary. Given this context, we seek to address the following question in this paper: how does the holography of information work in de Sitter space, where spatial slices have no boundaries?

With a view to addressing this question, we study expectation values of observables that act on the space of solutions of the Wheeler-DeWitt (WDW) equation recently found in \cite{dsfirst2023}. To begin with, this requires defining a norm on this space.  We propose a natural norm, obtained by integrating the square of the magnitude of the wavefunctional over field configurations and dividing by the volume of the group of diffeomorphisms and Weyl transformations. We show how this redundancy can be gauge-fixed using the Faddeev-Popov procedure \cite{Faddeev:1967fc,Faddeev:1973zb}.
 
Previously we showed \cite{dsfirst2023} that, in the nongravitational limit, the space of solutions to the WDW equation reduces to the space of dS invariant states defined by Higuchi using group averaging \cite{Higuchi:1991tk,Higuchi:1991tm,Marolf:2008it,Marolf:2008hg}. Higuchi defined a norm on this space by dividing the QFT norm of the states by the volume of the dS isometry group, resulting in a finite answer.  Here, we show that the  norm on the space of WDW solutions described above reduces to Higuchi's norm in the nongravitational limit. Our prescription also provides a systematic set of gravitational corrections to Higuchi's proposal.

Using our formalism, we turn to a  specific set of observables called  ``cosmological correlators''. These observables are physically significant and have attracted significant attention in the literature \cite{Maldacena:2002vr,Weinberg:2005vy,Arkani-Hamed:2018kmz,Baumann:2022jpr}.  They are usually expressed in terms of a product of local operators on the late-time slice of de Sitter space. Since the volume of the late-time slice is asymptotically large, these coordinates correspond to points that are separated infinitely in the physical spacetime and are global probes of the state. However, while such a product is a well-defined observable in a quantum field theory, it does not commute with the gravitational constraints. Hence, this description is not gauge invariant.

We propose that cosmological correlators should be understood as {\em gauge-fixed} observables. We provide a  prescription to compute the matrix elements of such observables between any two states of the theory. This set of matrix elements defines a gauge-invariant operator corresponding to every cosmological correlator. These operators are labelled by a set of coordinates on the late-time slice.
 
We show that our gauge-fixed observables are invariant under translations and rotations, and have simple transformation properties under scaling. Crucially, this property holds in all states of the theory, and not just in the Euclidean vacuum. 

Consequently, the specification of these observables in any open set $\reg$ suffices to specify them  everywhere. But the full set of cosmological correlators forms an overcomplete basis for all observables. Therefore, cosmological correlators in any arbitrary small part of the late-time Cauchy slice are sufficient to uniquely identify the state of the theory. 

Cosmological correlators can also be defined in quantum field theory. But in the absence of gravity, it is possible to construct states where they  coincide inside a small region but differ outside it. So the result above marks a sharp difference between the properties of gravitational and nongravitational theories. This provides the necessary generalization of the notion of holography of information to asymptotically de Sitter space.

Heuristically, this result can be put on the same footing as the results on the holography of information in AdS and in flat space.  There, the principle of holography of information implies that whenever a region $\reg$ is surrounded by its complement $\overline{\reg}$ then $\overline{\reg}$ contains all information about $\reg$.  This is simply because when spatial slices are noncompact, $\overline{\reg}$ extends to infinity and so it contains all information about the state.  In the present case, the spatial slices have the topology of $S^d$. Therefore every region $\reg$ both surrounds and is surrounded by its complement.  So it is natural for cosmological correlators in every region $\reg$ to have information about the entire state.

We present the holography of information in terms of  a precise mathematical result.  However this result should be interpreted with care. In particular, we do not suggest that a physical observer with access only to a small patch of spacetime can glean all information about the state using local measurements. 

First, as noted above, even the region $\reg$ is a small part of the late-time Cauchy slice, it still has infinite volume in the physical spacetime. Second, cosmological correlators are gauge-fixed observables that are merely labelled by a set of points. Since there are no local gauge-invariant observables in the theory, cosmological correlators also secretly correspond to nonlocal operators that cannot be measured through any strictly local process.

Moreover, in dS, it is not always fruitful to think in terms of external observers and so the question of what is physically observable might require us to construct a model of an observer who is part of the system.  Although, we do not seek to construct such a model in this paper, it is reasonable to envisage a model in which a physical observer can access low-point gauge-fixed observables of the kind we describe. But, as in AdS and in flat space, the identification of a sufficiently complicated state, $\reg$ from a small region requires very high-point cosmological correlators and presumably, in any reasonable physical model, such high-point correlators are effectively inaccessible.

An overview of this paper is as follows. In section \ref{secsummary}, we provide a summary of our results, including its key technical aspects. In section \ref{secnorm}, we discuss norms and expectation values in the space of solutions to the WDW equation. In section \ref{seccosmcorr}, we define cosmological correlators and study their properties. In section \ref{secholinfo}, we prove the principle of holography of information and discuss its implications. We conclude by discussing open questions in section \ref{secdiscussion}.

\section{Summary of results \label{secsummary}}
In a separate paper \cite{dsfirst2023}, we have shown that the space of solutions to the WDW equation with a positive cosmological constant, $\Lambda$, where the spatial slices have the topology of $S^d$  take on the asymptotic form
\be
\label{psiderstate}
\Psi[g, \chi] = e^{i S[g, \chi]}\sum_{n,m} \kappa^n \dcoeff[n,m]  Z_0[g,\chi]~.
\ee

This result involves several pieces of notation that we explain in turn. 
\begin{enumerate}

\item Here $g$ is the metric on a spatial slice and $\chi$ is a generic scalar matter field with scaling dimension $\Delta$. The solution is valid in the limit where the cosmological constant dominates the spatial curvature scalar $R$ (distinct from the spacetime curvature scalar), and other terms in the local energy density, everywhere on the slice. This requires the volume of the spatial slices to become asymptotically large compared to the cosmological scale. Physically, this corresponds to the late-time limit of an asymptotically de Sitter spacetime.

\item  The exponent $S[g,\chi]$ is a universal phase factor that comprises local functionals of $g$ and $\chi$ that diverge in the infinite volume limit, and was determined explicitly in \cite{dsfirst2023}.  $e^{i S[g,\chi]} Z_0[g,\chi]$  is the wavefunctional corresponding to the Euclidean vacuum, or the Hartle Hawking state \cite{Hartle:1983ai}. 

\item $Z_0[g,\chi]$ is invariant under diffeomorphisms and has the Weyl transformation property of a CFT partition function
\be
\label{weylinsummary}
\left(2  g_{ij} {\delta \over \delta g_{ij}} - \Delta \chi {\delta \over \delta \chi} \right) Z_0[g, \chi] = {\cal A}_d Z_0[g, \chi]~,
\ee
where ${\cal A}_d$ is an imaginary anomaly polynomial that is nonzero only in even $d$ and is determined explicitly in \cite{dsfirst2023}.
\item
The property above implies that, at the cost of a phase, it is possible to make a Weyl transformation to study $Z_0[g,\chi]$  in the vicinity of the flat metric, $g_{i j} = \delta_{i j} + \kappa h_{i j}$.\footnote{As explained in section 4.2 of \cite{dsfirst2023}, this Weyl transformation is made for convenience. To obtain the wavefunctional in the physical frame, where the metric describes a deformed sphere with large volume, one must use \eqref{weylinsummary} and obtain the correct phase. The phase factor will not appear in subsequent calculations in the main text but it is discussed further in Appendix \ref{wavephase}.}   In this Weyl frame, we can expand
\be
Z_0 = \exp[\sum_{n,m} \kappa^n \coeff[n,m]]~.
\ee
Here $\coeff[n,m]$ is a multilinear functional of the metric fluctuation $h_{i j}$ and the matter fluctuation $\chi$, 
\be
\label{coeffdef}
\coeff[n,m]\equiv {1 \over n! m!} \int d\vec{y}d\vec{z}\,h_{i_1 j_1}(y_1) \ldots h_{i_n j_n}(y_n) \chi(z_1) \ldots \chi(z_m) G_{n,m}^{\vI \vJ}(\vec{x}) ~.
\ee
As in \cite{dsfirst2023}, we vectorize the collective indices to condense the notation: $\vec{y}=(y_1, \ldots y_n), \vec{i} = i_1 \ldots i_n, \vec{j} = j_1 \ldots j_n, \vec{z} = z_1 \ldots z_m$ and $x$ is a generic coordinate, $\vec{x} = (\vec{y}, \vec{z})$. 
The wavefunction coefficients $G_{n,m}^{\vI\vJ}$ in \eqref{coeffdef} must obey a specific set of Ward identities analogous to those obeyed by correlators of a conformal field theory. 

\item Finally, $\dcoeff[n,m]$ is the {\em difference} of two distinct functionals, both of the form \eqref{coeffdef}, $\dcoeff[n,m] = \coeff[n,m] - \tcoeff[n,m]$. 

\end{enumerate}
In the nongravitational limit the sum over $n,m$ in \eqref{psiderstate} can be restricted to a single term. Away from this limit, the Ward identities link terms with different values of $n$.

It is conventional, in the cosmological literature to restrict attention to observables that only comprise field operators and do not contain insertions of their conjugate momenta. We will follow this convention in the main text of this paper. Such observables are sensitive only to $|\Psi[g,\chi]|^2$ and not to its phase. (See \cite{Maldacena:2015bha} for more discussion.)   However, from a mathematical perspective, there is no difficulty in generalizing our analysis to account for observables sensitive to the phase of the wavefunctional and we explain how to do this in Appendix \ref{wavephase}.

In this paper, we will propose that the natural norm on this space of wavefunctionals is obtained by simply squaring the asymptotic wavefunctionals, integrating over all field configurations and finally dividing by the volume of the \diffWeyl{}  group. More generally, the expectation value of a gauge-invariant operator $A$ is given by
\be
\label{expecAsummary}
(\Psi,A \Psi) =  {\cout\/\vdw}\int Dg D\chi\, \sum_{n,m,n',m'} \kappa^{n+n'} \dcoeff[n,m]^* \dcoeff[n',m'] |Z_0[g, \chi]|^2  A[g, \chi]  ~,
\ee
where $\cout$ is a physically unimportant normalization constant.

To parse this norm, we use a gauge-fixing condition which fixes the \diffWeyl{}  invariance. The gauge-fixing condition we choose is
\be\label{summgaugechoice}
\partial_{i} g_{i j} = 0; \qquad \delta^{i j} g_{i j} = d~.
\ee
The corresponding ghost action has zero modes that correspond to residual global symmetries that are not fixed by the gauge choice above.  

The zero modes correspond precisely to the generators of the conformal group in $d$-dimensions: translations, rotations, dilatations and special conformal transformations. For $d > 2$, the usual form of the special conformal transformations is corrected by a metric-dependent diffeomorphism. The integrated operators (inside $\dcoeff[n,m]$)   that appear in the correlator \eqref{expecAsummary} can be utilized to fix these residual symmetries. We fix three of the operators to
\be
\label{residualdg2}
x_1 = 0; \qquad x_2 = 1; \qquad x_3 = \infty~.
\ee
This choice, which is familiar from perturbative string theory, is enough to fix the residual conformal symmetry in all dimensions up to a residual $\sodminone$ invariance that is compact and can simply be excluded by hand or integrated over.

The notation $\overline{\dcoeff[n,m]}$ represents the operator obtained by fixing three of the points in an integrated product of operators like \eqref{coeffdef} using \eqref{residualdg2} with the appropriate measure factor. (See \eqref{overlinedef} for details.)
This leads to the gauge-fixed expression for the expectation value of an operator $A$ \be
(\Psi,A \Psi) = \sum_{n,m,n',m'}  \kappa^{n+n'} \big \langleQ \overline{\dcoeff[n,m]^* A[g, \chi] \dcoeff[n',m']} \big \rangleQ~,
\ee
where the symbol $\langleQ \,\cdot \,\rangleQ$ stands for
\be
\langleQ Q \rangleQ \equiv \cout \cfp  \int Dg D\chi \, \delta(g_{ii} - d) \delta(\partial_i g_{i j}) \Delta'_{\text{FP}}\, |Z_0[g, \chi]|^2 Q~.
\ee
Here, $\cfp$ is another physically irrelevant constant and $\Delta'_{\text{FP}}$ is a restricted Faddeev-Popov determinant obtained by integrating out the ghosts except for the zero modes.

\begin{figure}
\begin{center}
  \includegraphics[width=7cm]{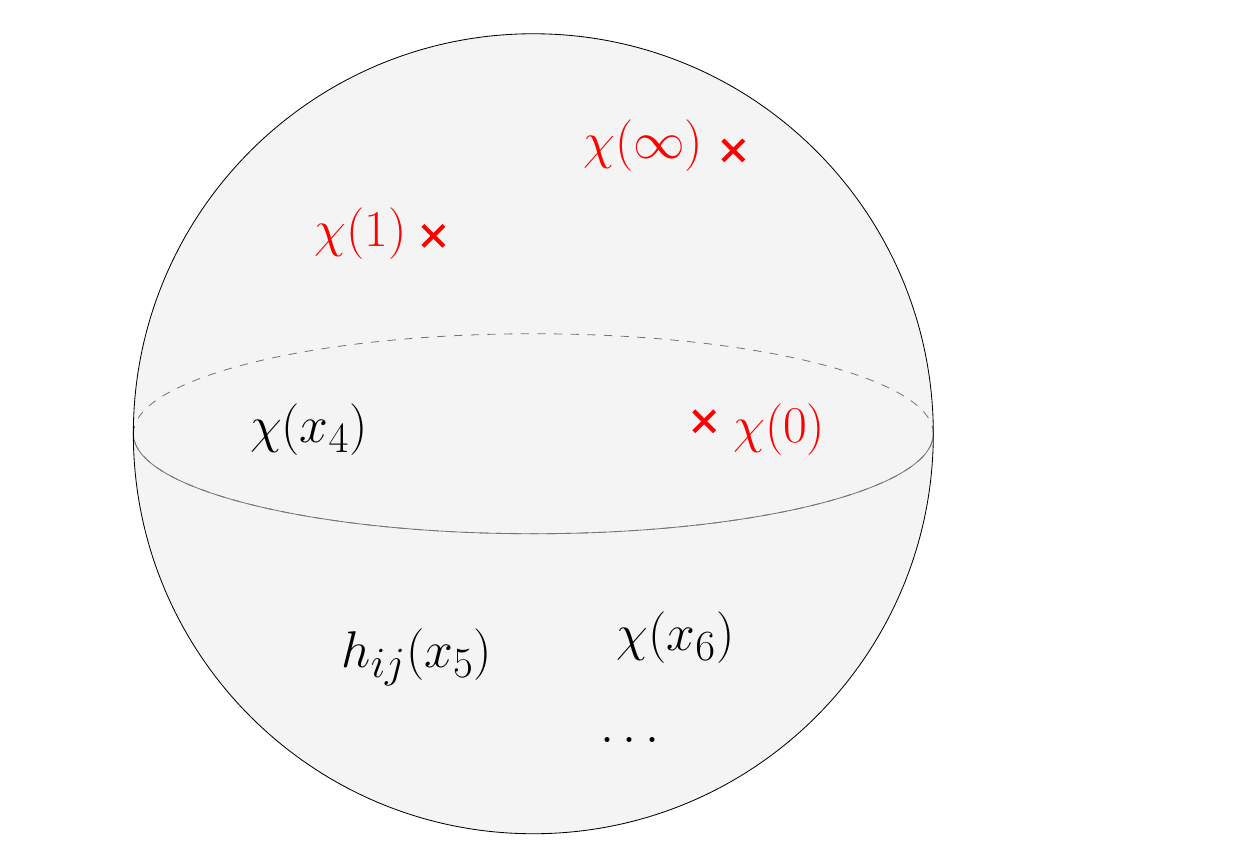}
					\caption{\em The residual gauge group is the Euclidean conformal group in $d$ dimensions $\r{SO}(1,d+1)$. Up to a compact subgroup, it can be fixed by fixing three points.  }
\end{center}
\end{figure}

At nonzero coupling the ghost determinant involves nontrivial factors of the metric. However, as $\kappa \rightarrow 0$ these factors vanish. In the nongravitational limit, the residual group can then also be handled by simply dropping the condition \eqref{residualdg2}, and instead dividing by the volume of the conformal group. The norm of a nongravitational state then becomes
\be
(\Psi_{\text{ng}}, \Psi_{\text{ng}}) = {\text{vol}(\sodminone) \over \text{vol}(\confgrp)} \lim_{\kappa \rightarrow 0} \,\langleQ \dcoeff[n,m]^* \dcoeff[n,m] \rangleQ~.
\ee
This is  precisely Higuchi's prescription for the norm: the RHS is the QFT norm  divided by the infinite volume of the conformal group. The factor of $\text{vol}(\sodminone)$ in the numerator arises due to a choice of normalization and is unimportant. Therefore our prescription leads to a derivation of Higuchi's proposal and also provides a precise prescription for how the norm should be generalized beyond $\k=0$.

Next, we turn to cosmological correlators. Cosmological correlators are labelled by points on the late-time slice of de Sitter space. While this makes sense in a quantum field theory, there are no local gauge-invariant observables in quantum gravity. We therefore propose that a cosmological correlator that is labelled by a product of $p$ insertions of the metric and $q$ insertions of the matter field, $\cprod^{p,q}_{\vI\vJ}(\vec{x})$,  (see \eqref{cprod} for notation) corresponds to a {\em gauge-fixed} observable:
\be
\label{cosmcorrsumm}
\llangle \Psi | \cprod^{p,q}_{\vI\vJ}(\vec{x}) | \Psi \rrangle_{\text{CC}}  \equiv \sum_{n,m,n',m'} \kappa^{n+n'} \langleQ \dcoeff[n,m]^* \dcoeff[n,m] \cprod^{p,q}_{\vI\vJ}(\vec{x}) \rangleQ~.
\ee
Note that the right hand side depends on the choice of gauge in \eqref{summgaugechoice} and also that the points in \eqref{cosmcorrsumm} have not been fixed by inserting delta functions for the residual gauge transformations and the corresponding zero-mode determinant but are simply fixed by hand.

The residual gauge transformations above turn into symmetries of cosmological correlators. Since special conformal transformations involve the metric fluctuation, they relate lower-point cosmological correlators to higher-point correlators. But we show that cosmological correlators are covariant under rotations, translations and dilatations in any state. Under translations and dilatations
\be
\llangle \Psi | \cprod^{p,q}_{\vI\vJ}(\lambda \vec{x} + \transv) | \Psi \rrangle_{\text{CC}} = \lambda^{-q \Delta} \llangle \Psi | \cprod^{p,q}_{\vI\vJ}(\vec{x}) | \Psi \rrangle_{\text{CC}} ~.
\ee

This leads us to a remarkable result: if one is given the cosmological correlators \eqref{cosmcorrsumm} in an arbitrarily small region, then this is sufficient to determine the correlators everywhere. But the set of correlators everywhere forms an overcomplete basis for the space of all observables. This means that for any region $\reg$
\be
 \llangle \Psi_1 | \cprod^{p,q}_{\vI\vJ}(\vec{x}) | \Psi_1 \rrangle_{\text{CC}} =  \llangle \Psi_2 | \cprod^{p,q}_{\vI\vJ}(\vec{x}) | \Psi_2 \rrangle_{\text{CC}}, ~~\forall \vec{x} \in \reg~\text{and}~\forall p,q \implies (\Psi_1, A \Psi_1) = (\Psi_2, A \Psi_2 ),
\ee
for any observable $A$.
This result provides the necessary generalization of the principle of holography of information to de Sitter space.

As mentioned above, the discussion of observables cosmological correlators is usually limited to products of the fields. But the set of cosmological correlators can be expanded to include the conjugate momenta as explained in Appendix \ref{wavephase}. The set of such generalized cosmological correlators in an arbitrarily small region suffices  to completely determine the full wavefunctional of the state, including its phase.

While this result marks a clear mathematical difference between quantum field theories and quantum gravity, it should be interpreted with caution. Cosmological correlators are secretly nonlocal observables. So the result above does not imply that a physical observer can determine the entire state of the universe through local measurements.

\section{Inner product and expectation values \label{secnorm} }

 In this section we discuss the problem of defining a norm on the space of solutions to the WDW equation that take the form \eqref{psiderstate}. We also show that in the nongravitational limit, this norm reduces to the norm defined by Higuchi. The definition of a norm also tells us how to compute expectation values of observables.

\subsection{The general problem}
We have determined the form of the wavefunctional in equation \eqref{psiderstate} only in the limit of large volume i.e. in the regime where the cosmological constant dominates the Ricci scalar of the spatial slice and the matter potential.  Nevertheless, we expect that this information is sufficient to define a norm on the Hilbert space.  The intuition is that the large-volume limit is equivalent to the late-time limit in the physical spacetime. In quantum mechanics, the norm of the state can be defined at any instant of time and does not require knowledge of the full time-evolution of the state. Therefore, we expect that the norm can be defined on the space of wavefunctionals in the large-volume limit and should not require details of the wavefunctional everywhere in the configuration space.

Once the question has been reduced to that of finding the norm on states of the form \eqref{psiderstate}, we find another simplification. Although the wavefunctional $\Psi$ itself has a phase factor that is not Weyl invariant, and $Z[g,\chi]$ might have a Weyl anomaly, $|\Psi|^2$ is \diffWeyl{}   invariant since the phase factor cancels and the anomaly is pure imaginary. So it makes sense to study $|\Psi|^2$ beyond the domain of large-volume metrics where the form \eqref{psiderstate} was originally derived. (This point is discussed in some more detail in section 4.2 of \cite{dsfirst2023}.)  

We propose that the norm of a wavefunctional $\Psi$ is given by considering the integral of $|\Psi|^2$ over all field configurations and dividing by the volume of the  group of diffeomorphisms and Weyl transformations.
\be
\label{normproposal}
(\Psi,\Psi) \equiv { \cout \/\vdw}\int Dg D\chi\, \sum_{n,m,n',m'} \kappa^{n+n'} \dcoeff[n,m]^* \dcoeff[n',m'] |Z_0[g, \chi]|^2~.
\ee
Here $\cout$ is an overall state-independent normalization constant that we will choose below for convenience. 

Now consider a \diffWeyl{} invariant operator $A[g, \chi]$ that maps states of the form \eqref{psiderstate} back to the state space. We propose that the expectation value of the operator is given by
\be
\label{expecA}
(\Psi,A \Psi) =  {\cout\/\vdw}\int Dg D\chi\, \sum_{n,m,n',m'} \kappa^{n+n'} \dcoeff[n,m]^* \dcoeff[n',m']  |Z_0[g, \chi]|^2 A[g, \chi]  ~.
\ee
Note that the knowledge of the norm for the state  $(a |\Psi_1 \rangle + b |\Psi_2 \rangle)$, and the expectation value of $A$ in this state, for all $a$ and $b$ is sufficient to determine the overlap $(\Psi_1, \Psi_2)$ and the matrix elements $(\Psi_1, A  \Psi_2)$ including their phase.

The proposal for the norm and expectation value, \eqref{normproposal} and \eqref{expecA}, is not unique but we adopt it because it is natural and simple. It might be of interest to explore alternative norms, as we briefly discuss in section \ref{secsubtlenorm}. We also postpone a discussion of some subtle aspects of the proposal  to section \ref{secsubtlenorm}. For now, we proceed to examine the technical problem of gauge fixing the \diffWeyl{} redundancy to obtain a practical method of computing the norm. In the section below, we use the Faddeev-Popov formalism to obtain a gauge-fixed expression.  In Appendix \ref{appbrst}, we show that the gauge-fixed functional integral is invariant under a BRST transformation.

\subsection{Gauge-fixing conditions}

In order to implement the Faddeev-Popov procedure to gauge fix the functional integral, we use the following gauge-fixing conditions
\be\label{GFconds}
\p_i g_{ij}=0; \qquad g_{ii}=d~.
\ee
We use the standard summation convention, so that repeated indices are summed over. The derivative that appears in \eqref{GFconds} is an {\em ordinary} partial derivative and so the gauge-fixing condition explicitly breaks both diffeomorphism invariance and Weyl invariance. With $g_{ij} = \delta_{ij}+\kappa h_{ij}$, our choice requires $h_{ij}$ to be traceless and transverse. 

In $d=2$, the conditions \eqref{GFconds} are equivalent to fixing $g_{i j}$ to $\delta_{i j}$. However, for $d > 2$ it is, in general, not possible to fix the metric to a ``fiducial metric'' using only diffeomorphisms and Weyl transformations.

We adopt the gauge choice \eqref{GFconds} for simplicity. In Appendix \ref{app:residual}, we discuss alternate choices of gauge that lead to the same physical results. 

The infinitesimal variation due to a diffeomorphism $x^i \to x^i + \xi^i$ and a Weyl transformation $g_{ij}\to e^{2\varphi}g_{ij}$ of the metric is given by
\be
\d_{(\xi,\vphi)}g_{ij}
=  \n_i  \xi_j +  \n_j \xi_i +2\vphi g_{ij}~,
\ee
where $\xi_i = g_{i k} \xi^k$.
It will be convenient below to change the parameter of the Weyl transformation to implement the shift  $\varphi \to \varphi-\frac{1}{d} \nabla_k \xi_k$.
The infinitesimal  transformation now takes the form
\be
\d_{(\xi,\vphi)}g_{ij}=  (P \xi)_{i j} +2\vphi g_{ij}~,
\ee
where we have defined
\be
\label{pdef}
\begin{split}
(P \xi)_{i j} &\equiv g_{j k} \n_i \xi^k+ g_{i k} \n_j\xi^k-\frac{2}{d}g_{ij}\nabla_k \xi_k \\ &=\xi^\ell \partial_\ell g_{ij}+g_{jk}\partial_i\xi^k  +   g_{ik}\partial_j \xi^k  -\frac{2}{d}g_{ij }g_{k\ell} \partial_{\ell}\xi^k~.
\end{split}
\ee
The shift is chosen so that the $(P \xi)_{i j}$ is traceless provided $g_{ii} = d$.  

\subsubsection{Residual gauge transformations}
The gauge fixing conditions \eqref{GFconds} do not completely fix the gauge. Since $(P \xi)_{i j}$ is traceless provided $g_{ii} = d$, the residual symmetry corresponds to solutions of the equation
\be
\label{residualsolution}
\enc{\mathcal{D} \xi}_j \equiv  \p_i (P \xi)_{i j} = 0~.
\ee
Solutions of this equation are in one-to-one correspondence with the generators of ${\confgrp}$. However, the nature of the solutions is slightly different for $d > 2$ and for $d=2$.

It is shown in Appendix \ref{app:residual} that, for a general metric, in $d > 2$,  there are  ${(d+1) (d+2) \over 2}$ solutions of \eqref{residualsolution}. These are given by
\be
\label{residualgrav}
\begin{split}
\text{translations}&:\quad \xi^i  =\a^i; \\
\text{rotations}&:\quad \xi^i =M^{i j}  x^j\\
\text{dilatations}&:\quad \xi^i  = \lambda x^i  \\
\text{SCTs} &: \quad \xi^i  = ( 2 (\beta \cdot x) x^i  - x^2 \beta^i ) +  \beta^j v^{i}_{j}
\end{split} 
\ee
where $\lambda$, and $M^{ij}$ denote, respectively, a number and an antisymmetric matrix and $\a^i$ and $\beta^i$ are vectors. 

The notable aspect of \eqref{residualgrav} is that the usual special conformal transformations are corrected as noted in \cite{Hinterbichler:2013dpa,Ghosh:2014kba}.  The matrix $v^{i}_{j}$ depends nontrivially on the metric and vanishes when $g_{i j} = \delta_{i j}$.  In Appendix \ref{app:residual}, we present an algorithm to find $v^{i}_{j}$ in perturbation theory.  It is also shown there that although the SCT itself is modified, the algebraic structure of the residual transformations \eqref{residualgrav} remains that of $\confgrp$. Appendix \ref{app:residual} also discusses residual gauge transformations for other choices of gauge.

In $d=2$, since the conditions \eqref{GFconds} fix $g_{i j} = \delta_{i j}$, the correction term in the SCT always vanishes. 
\be
 v^{i}_{j} = 0, \qquad \text{for}~d=2~.
\ee
Appropriate linear combinations of the two allowed SCTs in $d=2$ correspond to the two independent special conformal transformations that are usually described in terms of ``holomorphic'' and ``anti-holomorphic'' transformations in the discussion of string perturbation theory.

\subsubsection{Fixing the residual symmetry}
To fix the residual gauge symmetry, we will take advantage of the presence of insertions in \eqref{expecA}. We will assume that the state under consideration has at least two insertions, which implies the presence of at least four insertions in \eqref{expecA}. In all dimensions, the residual gauge symmetry can then be fixed by setting the position of three insertions as follows:
\be
\label{gaugeFresd2}
x_1 = 0; \qquad x_2 = 1; \qquad x_3 = \infty~.
\ee
The choice of a point at the origin and another point at infinity fixes the translations and special conformal transformations. Fixing $x_2$ to $1 \equiv (1,0,\dots,0)$ fixes the dilatations and also part of the rotations.  This choice does not fix the $\sodminone$ group of rotations of the hyperplane orthogonal to the $0-1$ axis. But since this group is compact, it can simply be integrated over and does not lead to any divergence in the functional integral.  It is convenient to impose the last condition using the coordinates $\inv{x}_3^i = {x_3^i \over |x_3|^2}$ so that it can be written as $\inv{x}_3 = 0$.

\subsection{Faddeev-Popov procedure}
To gauge fix the functional integral for the expectation value of an operator in \eqref{expecA}, we insert the following expression for the identity,
\be \label{insertionone}
1 = \Delta_\text{FP}  \int  D\xi D\vphi\,\delta(g_{ii}^{(\xi,\vphi)}-d)\delta (\p_i g_{ij}^{(\xi,\vphi)}) \delta(x_1)  \delta(x_2 - 1) \delta(\inv{x}_3)~,
\ee
where the notation $g^{(\xi, \vphi)}$ indicates the metric obtained upon acting on $g_{i j}$ with the diffeomorphism parameterized by $\xi$ and the Weyl transformation $\vphi$. $\Delta_\text{FP} $ is the standard Faddeev Popov determinant that we will evaluate below. 

Substituting the infinitesimal transformations in \eqref{insertionone},  we can write
\be \label{kek}
\begin{split}
\Delta_\text{FP}^{-1} = \int  D\xi D\vphi\,\delta(2 d \vphi)\, \delta ( (\mathcal{D} \xi)_{i}) \,   \delta(\xi^j(0)) \, \delta(\xi^j(1)) \delta(\inv{\xi}^j(\infz)) ~,
\end{split}
\ee
where, at infinity, we use the diffeomorphism in the inverted chart
\be
\inv{\xi}^i(x) = {1 \over |x|^2} \Big(\xi^{i}(x)- 2 (x\cdot \xi) {x^{i} \over |x|^2} \Big)~,
\ee
which is inserted at $x=\infty$ corresponding to $\tilde{x}=0$. The delta function for $\vphi$ is trivial, and one can simply integrate it out. 

The Faddeev-Popov determinant may be evaluated using the standard procedure of first writing the delta functions as integrals over auxiliary parameters, and then simply replacing the bosonic parameters by Grassmann numbers. 
This leads to an expression for $\Delta_{\text{FP}} $ in terms of a  $c$-$\bar{c}$ ghost action:
\be
\label{deltafp}
\Delta_{\text{FP}} =  \cfp \int D c D \bar{c} \,e^{-S_\ghosts }  \big(\prod_{j} c^j(0) c^j(1) \inv{c}^j(\infz) \big)~,
\ee
where the $c$ ghost insertions correspond to $\xi$ insertions in \eqref{kek} and the ghost action (derived in \cref{appbrst}) $S_\ghosts $ is given by
\be
\label{ghostaction}
S_\ghosts = \int d^d x \,\bar{c}^{j} ({\mathcal D} c)_{j} ~.
\ee

The ghost action \eqref{deltafp} has zero modes corresponding to the residual gauge transformations discussed previously. Some of these are soaked up by the insertion of the $3 d$ c-ghosts in the denominator. But in the ghost functional-integral \eqref{deltafp}, we  {\em exclude} the zero modes that correspond to rotations that leave the point  $x_2 = 1$ invariant. (All rotations leave the origin and the point at $\infty$ invariant.)  These zero modes correspond to the unfixed compact part of the residual symmetries and if we were to integrate over them we would obtain zero since there is nothing to soak them up. But there is no difficulty in excluding them in the functional integral since they are orthogonal to all other modes. These unfixed residual transformations also contribute a factor of $\text{vol}(\sodminone)^{-1}$ in $\Delta_{\text{FP}}$ but this can be absorbed in the overall normalization constant $\cfp$. 

We do not keep track of the overall constant $\cfp$.  This factor always drops out of any physical computation since the same constant appears in both the norm and the expectation value and so $(\Psi, \Psi)^{-1} (\Psi, A \Psi)$ does not depend on this constant.

Combining everything together, the  gauge-fixed expression for the expectation value of $A$ can be written in the following form.
\be
\label{finalfixedexpecA}
\begin{split}
(\Psi,A \Psi) &=   \cout \cfp \int Dg D\chi\,  D c D \bar{c} \, \sum_{n,m,n',m'} \kappa^{n+n'} \dcoeff[n,m]^* A[g, \chi] \dcoeff[n',m'] |Z_0[g, \chi]|^2  e^{- S_\ghosts } \\
&\times  \delta(g_{ii}-d)\delta (\p_i g_{ij}) \delta(x_1) \delta(x_2-1)\delta(\inv{x}_3) \big(\prod_{i} c^i(0) c^i(1) \inv{c}^i(\infz) \big)~.
\end{split}
\ee
It is understood that the points $x_1, x_2, x_3$  correspond to operators that are part of $A$ or $\dcoeff[n,m]$.

In Appendix \ref{appbrst} we show that the gauge-fixed integral \eqref{finalfixedexpecA} remains invariant under a BRST symmetry when the delta functions are implemented using auxiliary fields.

\paragraph{\bf Ghost determinant.}
The expression \eqref{finalfixedexpecA} can simplified by evaluating the ghost determinant. First we expand the $c$-ghosts using a basis of orthonormal vector fields. The correct inner product between vector fields is the one on the sphere. (See Appendix \ref{app:residual} for more discussion.)

We then divide the space of vector fields into the subspace of zero modes and the subspace of nonzero modes. Since we have excluded modes corresponding to rotations that leave $(1,0,\dots,0)$ invariant, the remaining subspace of zero modes is exactly $3 d$ dimensional.  Using the index $z$ to run over zero modes and the index $n$ to run over the non-zero modes, we can write
\be
c^j = \sum_{z} c_{(z)} \zmv_{(z)}^j + \sum_{n} c_{(n)} \zmv_{(n)}^j ~.
\ee
First, consider the contribution of the non-zero modes.  This can be evaluated by neglecting any $c$ insertions outside the ghost action. This is because in the ghost action, the nonzero modes of $c$ are always paired with a mode of $\bar{c}$. Upon series expanding the action, further $c$ insertions simply give zero either in the integral over the $c$ modes or the $\bar{c}$ modes (for further details, see \cite{Polchinski:1998rq}). Then, to obtain the non-zero mode contribution, we simply perform the integral over the ghost action to obtain a restricted FP determinant
\be
\label{fprestricted}
\int D \bar{c} D c'  \, e^{-S_\ghosts} = \Delta_{\text{FP}}'~,
\ee
where the prime label indicates that the zero modes have been excluded from the measure. Note that the above notation is somewhat deceptively compact since this restricted determinant depends on the metric fluctuation.

We now turn to the zero mode contribution. The zero-mode fields are proportional to those given in \eqref{residualgrav} but we will fix the normalization below for convenience.  We can choose $d$ modes to correspond to translations in the $d$-possible directions; one mode corresponds to dilatations; $d$ modes correspond to special conformal transformations; and $(d-1)$ modes correspond to rotations with $M^{i j} \propto \delta^{i}_{i_0} \delta^{j}_1 - \delta^{i}_{1} \delta^{j}_{i_0}$ with $i_0 \neq 1$. The index $z$ runs over all these $3 d$ fields and we can therefore construct the $3 d \times 3 d$ matrix
\be
M = \begin{pmatrix}
 \zmv_{(1)}^{1}(0) & \dots &
 \zmv_{(1)}^{d}(0) & \zmv_{(1)}^{1}(1) & \dots &
 \zmv_{(1)}^{d}(1) &\inv{\zmv}_{(1)}^{1}(\infty) & \dots &
 \inv{\zmv}_{(1)}^{d}(\infty) \\
 \zmv_{(2)}^{1}(0) & \dots &
 \zmv_{(2)}^{d}(0) & \zmv_{(2)}^{1}(1) & \dots &
 \zmv_{(2)}^{d}(1) &\inv{\zmv}_{(2)}^{1}(\infty) & \dots &
 \inv{\zmv}_{(2)}^{d}(\infty) \\
 \vdots  & \dots &\vdots  &  \vdots  & \dots &\vdots  &
\vdots  & \dots &
\vdots  \\
 \zmv_{(3d)}^{1}(0) & \dots &
 \zmv_{(3d)}^{d}(0) &\zmv_{(3d)}^{1}(1) & \dots &
 \zmv_{(3d)}^{d}(1) &\inv{\zmv}_{(3d)}^{1}(\infty) & \dots &
 \inv{\zmv}_{(3d)}^{d}(\infty) \\
\end{pmatrix}~.
\ee
The zero-mode determinant is
\be
\Delta_{\text{FP}}^0 = \text{det}(M)~.
\ee
We now find that our gauge choice leads to a simplification. The special conformal transformations depend on the metric through $\resv^{i}_{j}$ as shown in \eqref{residualgrav}. However, this dependence vanishes at infinity. Moreover, while the special conformal transformations become a constant at infinity, all other zero-mode fields vanish at infinity. Therefore $\text{det}(M)$ does not depend on the special conformal transformations at the points $0$ or $1$ and thus $\text{det}(M)$ is independent of the metric. By normalizing the zero-mode fields appropriately, we can simply set
\be
\Delta_{\text{FP}}^{0} = 1~.
\ee

\paragraph{\bf Final answer.}
We now introduce some notation and present our final answer in a compact form.  When three points within an integrated product of operators are fixed using the delta functions that fix the residual transformations, we denote this using a overline. For instance,
\be
\label{overlinedef}
\overline{\dcoeff[n,m]} \equiv \int d \vec{x}\, \delta(x_1) \delta(x_2 - 1) \delta(\inv{x}_3) G_{n,m}^{\vI\vJ}(\vec{x}) h_{i_1,j_1}(x_1) h_{i_2,j_2}(x_2) \ldots \chi(x_{n+1}) \ldots \chi(x_{n+m})~.
\ee
The notation $\overline{\dcoeff[n,m] A[g,\chi] \dcoeff[n,m]^*}$ allows for the position of any three operators in the product to be fixed.

Next, in the expression for the functional integral \eqref{finalfixedexpecA}, we choose
\be
\cout = {1 \over \cfp} \left[\int Dg D\chi \,  \delta(g_{ii}-d)\delta (\p_i g_{ij}) \Delta'_{\text{FP}}|Z_0[g, \chi]|^2\right]^{-1}~.
\ee 
This choice makes the product $\cout \cfp$ equal to the inverse of the functional integral over the wavefunctional of the Euclidean vacuum. Hence we should think of physical observables as the \textit{ratio} of a functional integral with operator insertions, and the functional integral over the Euclidean vacuum.

Given a general product of the metric and other matter fields, $Q$, we also define the notation 
\be
\label{expecQ}
\langleQ Q \rangleQ = \cout \cfp  \int Dg D\chi \,\delta(g_{ii} - d) \delta(\partial_i g_{i j})\Delta'_{\text{FP}} |Z_0[g, \chi]|^2 Q  . 
\ee
Intuitively, the notation can be thought of as the expectation value of $Q$ in the Euclidean vacuum although this intuition should be used with care since (see section \ref{secsubtlenorm}) the vacuum itself might not be normalizable.

Using this notation, we can then rewrite the gauge-fixed path integral \eqref{finalfixedexpecA} as
\be
\label{compactfinalfixedA}
(\Psi,A \Psi) = \sum_{n,m,n',m'}  \kappa^{n+n'} \big \langleQ \overline{\dcoeff[n,m]^* A[g, \chi] \dcoeff[n',m']} \big \rangleQ~.
\ee
Note that setting $A = 1$ yields the norm.
\be
\label{compactfinalnorm}
(\Psi, \Psi) = \sum_{n,m,n',m'}  \kappa^{n+n'} \big \langleQ \overline{\dcoeff[n,m]^* \, \dcoeff[n',m']} \big \rangleQ~.
\ee
The relations \eqref{compactfinalnorm} and \eqref{compactfinalfixedA} represent our final answers in compact form.

\subsection{Nongravitational limit \label{subsecnormnongrav}}

We now show that our expression for the norm coincides precisely with the norm proposed by Higuchi \cite{Higuchi:1991tk,Higuchi:1991tm} in the nongravitational limit.

It was explained in \cite{dsfirst2023} that the form of the allowed states simplifies in the nongravitational limit.  More specifically, in the nongravitational limit, with the state corresponding to the Euclidean vacuum denoted by $| 0 \rangle$ the allowed states take the form
\be
\label{freestates}
|\Psi_{\text{ng}} \rangle = \int d \vec{x} \, \delta G_{n,m}^{\vI\vJ}(\vec{y}, \vec{z})  h_{i_1 j_1}(y_1) \ldots h_{i_n j_n}(y_n) \chi(z_1) \ldots \chi(z_m) | 0 \rangle~.
\ee
The simplification above is that we do not have a sum over multiple values of $n$ that is necessary when $\kappa \neq 0$ by the Ward identities.

Now consider the nongravitational limit of the  expectation value defined in \eqref{expecQ},
\be
\label{expecqng}
\langle 0 |  Q  | 0 \rangle_{\text{QFT}} \equiv \lim_{\kappa \rightarrow 0}\, \langleQ Q \rangleQ  \,~.
\ee
In the nongravitational limit,  the ghosts decouple from the metric. 
Since $\Delta_{\text{FP}}'$ has no dependence on the metric fluctuation in the limit $\kappa \rightarrow 0$, it trivializes to a numerical factor. The gauge conditions still ensure that $h_{i j}$ is transverse and traceless. Therefore, this expression instructs us to integrate the product of insertions over the matter fields and over the transverse traceless fluctuations of the metric using the $\kappa \rightarrow 0$ limit of the wavefunctional for the Euclidean vacuum. This is precisely how one would have computed the expectation value of the product of operators in the quantum field theory, including the fluctuations of free transverse-traceless gravitons. This explains the choice of notation in \eqref{expecqng}.

Now consider the norm of two states of the form \eqref{freestates}.  Our final answer for the norm in the nongravitational limit can be written as
\be
\label{ourngnorm}
\begin{split}
& ( \Psi_{\text{ng}} ,\Psi_{\text{ng}} ) = \int d \vec{x} d \vec{x}' \, \delta G^{\vI \vJ}_{n,m}(\vec{x})  \, (\delta G^{\vI \vJ}_{n,m})^*(\vec{x}') \delta(x_1) \delta(x_2 - 1) \delta(\inv{x}_3) \\ &\times \langle 0 | h_{i_1' j_1'}(y'_1) \ldots h_{i'_n j'_n}(y'_n) \chi(z'_1) \ldots \chi(z'_m)  h_{i_1 j_1}(y_1) \ldots h_{i_n j_n}(y_n) \chi(z_1) \ldots \chi(z_m) | 0 \rangle_{\text{QFT}}
\end{split}
\ee
where $x_1, x_2, x_3$ can be any three coordinates from the $\vec{x}=(\vec{y}, \vec{z})$ or $\vec{x}'=(\vec{y}', \vec{z}')$ that appear above.

We recognize that this is just the gauge-fixed version of Higuchi's proposal as we can undo the residual gauge-fixing and write this as a group average which becomes
\be
\label{higuchingnorm}
\begin{split}
& ( \Psi_{\text{ng}} ,\Psi_{\text{ng}} ) = {\text{vol}(\sodminone) \over \text{vol}(\confgrp)}  \int d \vec{x} d \vec{x}' \, \delta G^{\vI \vJ}_{n,m}(\vec{x})  \, (\delta G^{\vI \vJ}_{n,m})^*(\vec{x}') \\ &\times \langle 0 | h_{i_1' j_1'}(y'_1) \ldots h_{i'_n j'_n}(y'_n) \chi(z'_1) \ldots \chi(z'_m)  h_{i_1 j_1}(y_1) \ldots h_{i_n j_n}(y_n) \chi(z_1) \ldots \chi(z_m) | 0 \rangle_{\text{QFT}}~.
\end{split}
\ee
This can be derived  repeating the steps in section 5.3 of \cite{dsfirst2023}. We can also write this as
\be
( \Psi_{\text{ng}} ,\Psi_{\text{ng}} ) ={\text{vol}(\sodminone) \over \text{vol}(\confgrp)} \ln\Psi_\r{ng}|\Psi_\r{ng}\rn_\r{QFT}
\ee
and we recognize Higuchi's inner product. In this expression the infinite volume of the conformal group in the denominator cancels the infinite QFT norm. The additional finite factor of $\text{vol}(\sodminone)$ emerges because an $\sodminone$ subgroup of $\confgrp$ leaves three points invariant. Since this is an overall finite normalization constant in the norm, it is physically irrelevant.

Note that it would {\em not} be correct to equate \eqref{ourngnorm} with \eqref{higuchingnorm} away from the nongravitational limit even after replacing the QFT expectation values with a gravitational expectation value of the form \eqref{expecQ}. First, away from this limit the form of the states shown in \eqref{freestates} is corrected. More importantly the action of a special conformal transformation on the operators that appear there is corrected due to the correction term in \eqref{residualgrav}. Consequently, special conformal transformations relate an expectation value to another expectation value with additional metric insertions. Therefore, away from the gravitational limit the gauge-fixed integrand that appears in \eqref{ourngnorm} cannot simply be equated with a group average.

Therefore our proposal \eqref{compactfinalnorm} reduces to Higuchi's proposal in the nongravitational limit but also provides a systematic method of correcting it at nonzero $\kappa$.

If, in addition to the nongravitational limit,  we consider the free-field limit for matter fields in the principal series then the space of ``conformal blocks'' naturally provides an orthonormal basis for the Hilbert space under the norm \eqref{higuchingnorm}. This interesting point is discussed further in Appendix \ref{appconformal}.

\subsection{Subtleties \label{secsubtlenorm}}
In the technical discussion of the norm, we have glossed over some subtleties that we now list. 
 \begin{enumerate}
 \item
\label{commentmeasure}
In even dimensions, the transformation of the measure might introduce  a Weyl anomaly in \eqref{compactfinalfixedA}  and \eqref{compactfinalnorm} \cite{Fujikawa:1980vr}. Relatedly, in string theory, where a similar functional integral appears, the critical dimension is fixed by demanding that the Weyl anomaly vanishes.  So, in even $d$ the expression for the norm might need to be improved by adding auxiliary fields to preserve \diffWeyl{} invariance.  However, we also note that we have some more freedom because $|Z_0[g, \chi]|^2$ is not a local functional and therefore it might be reasonable to study nonlocal measures. We leave a deeper study of the measure to future work. For odd $d$, which includes the case $d = 3$ of physical interest since it corresponds to an asymptotically dS$_4$ spacetime,  we do not expect these issues to arise.
\item \label{nonnormhh}
The question of the normalizability of the Hartle-Hawking wavefunctional and its relation to the nonperturbative instability of de Sitter space has been discussed in \cite{Anninos:2012ft,Anninos:2013rza,Castro:2012gc}.  This is related to the question of the measure.  We will not address this issue in this paper. We note that physical quantities are always related to the {\em ratio}  of  an expression of the form \eqref{compactfinalfixedA} and an expression of the form \eqref{compactfinalnorm} ,  which might be better behaved. 
\item
The formula \eqref{compactfinalfixedA} requires the presence of at least three operators in the product $\dcoeff[n,m]^* A[g, \chi] \dcoeff[n',m']$. Therefore it cannot be used to compute the norm of the original Euclidean vacuum. (This problem is separate from the one discussed in point \ref{nonnormhh}.) This suggests that the vacuum state itself is not part of the Hilbert space at all and only excitations above the vacuum are normalizable states. This issue was noted earlier by Higuchi \cite{Higuchi:1991tm} and also, recently, in \cite{Chandrasekaran:2022cip}. It is similar to the one that arises in string perturbation theory, if one attempts to define the sphere partition function with less than three vertex operator insertions. It would be nice to understand this better, perhaps using the techniques of \cite{Erbin:2019uiz,Mahajan:2021nsd,Anninos:2021ene, Eberhardt:2021ynh,Ahmadain:2022tew}.
\item
\label{threeremovediv}
Consider a term with a given value of $n,m$ in the expression \eqref{compactfinalfixedA}. This involves the ``expectation value'' of a product of operators integrated with coefficient functions that are conformally covariant. (Recall the definition of $\coeff[n,m]$ in \eqref{coeffdef}.) It will be shown in section \ref{seccosmcorr}, that the expectation value also transforms in a simple fashion under the conformal group.  When combined with the coefficient function this produces an integrand that is invariant under rotations, dilatations and translations and in the $\kappa \to 0$ limit under SCTs. Fixing three points in such an integral suffices to remove an obvious divergence that comes from the volume of the conformal group.  
\item
Nevertheless,
there might be additional divergences in \eqref{finalfixedexpecA} that arise due to the ``collision'' of operators. This issue again parallels an issue that appears in string perturbation theory. We hope that the ideas developed to deal with these divergences in that setting, including a suitable $i \epsilon$ prescription \cite{Witten:2013pra},  the use of string field theory techniques \cite{Sen:2014dqa} and off-shell methods \cite{Fradkin:1985ys,Fradkin:1984pq,Ahmadain:2022tew}, will be effective in this setting as well. We leave further study of this issue to future work.
\end{enumerate}

\section{Cosmological correlators \label{seccosmcorr}}
Cosmological correlators are of interest since they provide a leading-order approximation to the fluctuations generated during the inflationary epoch, when the universe could be approximated by a de Sitter spacetime. In this section, we will define these quantities within our framework and discuss some of their properties.

\subsection{Definition of cosmological correlators}
In the literature, cosmological correlators are usually computed as QFT-expectation values of the form $\langle \chi(x_1) \ldots \chi(x_n) \rangle_{\text{QFT}}$, where $x_i$ are points on the late-time boundary of de  Sitter. Note that since these points are on the asymptotic late-time slice, they are infinitely separated in physical distance for any finite separation of the coordinates. In a quantum field theory, the meaning of such correlators is clear. However, in a theory of quantum gravity, the product of local operators on the late-time slice does not commute with the constraints and so is not gauge invariant.

For instance,  under a diffeomorphism $x^i \rightarrow x^i + \xi^i$, an operator insertion $\chi(x)$ transforms as
\be
\chi(x) \rightarrow \chi(x) + \xi^i \partial_{i} \chi(x)~,
\ee
and thus does not remain invariant. Since diffeomorphisms on the late-time slice are generated by the momentum constraint, this means that the operator $\chi(x)$ does not commute with the momentum constraint. Likewise, it may be checked that the operator does not commute with the Hamiltonian constraint. This is expected from the well-known result \cite{DeWitt:1967yk} that gauge-invariant observables in gravity cannot be local.

Nevertheless, it is possible to make sense of such operators by fixing the gauge. We propose the following definition of cosmological correlators. Let 
\be
\label{cprod}
\cprod^{p,q}_{\vI\vJ}(\vec{x}) = h_{i_1 j_1}(z_1) \ldots h_{i_p j_p}(z_p) \chi(y_1) \ldots \chi(y_q)~,
\ee
denote a product of $p$ metric fluctuations and $q$ matter fluctuations. 
Now consider a state of the form \eqref{psiderstate}. We propose that the cosmological correlator corresponding to the product \eqref{cprod} in the state \eqref{psiderstate} be defined as
\be
\begin{split}
&\llangle \Psi | \cprod^{p,q}_{\vI\vJ}(\vec{x}) | \Psi \rrangle_{\text{CC}}= \sum_{n,m,n',m'} \kappa^{n+n'} \langleQ \dcoeff[n,m]^* \dcoeff[n,m] \cprod^{p,q}_{\vI\vJ}(\vec{x}) \rangleQ~,
\end{split}
\ee
using the expectation value \eqref{expecQ}. This can be written more explicitly as 
\be
\label{ccprescription}
\llangle \Psi | \cprod^{p,q}_{\vI\vJ}(\vec{x}) | \Psi \rrangle_{\text{CC}}  \equiv  \cout \cfp \int D g D \chi D \bar{c} D c' e^{-{\mathcal{S}}} \cprod^{p,q}_{\vI\vJ}(\vec{x}) ~,
\ee
where, for convenience in the discussion below,  we have introduced an ``action'' $\cS$ 
\be
e^{-\mathcal{S}} \equiv  e^{-S_\ghosts}  \delta(g_{ii} - d) \delta(\partial_i g_{i j})   |Z_0[g, \chi]|^2  \sum_{n,m,n',m'} \kappa^{n+n'} \dcoeff[n,m]^* \dcoeff[n,m] ~.
\ee

Let us discuss some features of our proposed correlator.

\begin{enumerate}

\item Recalling point \ref{threeremovediv} in section \ref{secsubtlenorm}, our prescription for the correlator makes sense provided that the product \eqref{cprod} has at least three points.  $\dcoeff[n,m]$ also contains products of the form  \eqref{cprod} integrated with conformally covariant functions. Therefore, if we study a $k=p+q$-point cosmological correlator, each term in the sum in \eqref{ccprescription} is an expectation value of a product of $(n+m)+(n'+m')+k$ operators where $(n+m)+(n'+m')$ operators are integrated with a conformally covariant function. It will be shown below that the expectation value is conformally covariant. (See section \ref{subsecsymm} for a precise discussion.) Therefore a value of $k \geq 3$ is sufficient to remove a potential divergence from the volume of the conformal group.  It would be nice to understand two-point correlators, perhaps, by  generalizing the methods of  \cite{Erbin:2019uiz}.

\item The prescription \eqref{ccprescription} continues to make sense if we remove the insertions of $\dcoeff[n,m]$ and consider only the vacuum state. In the vacuum state, the restriction $k \geq 3$ does not apply. Although the vacuum is the state that is most commonly used to compute cosmological correlators, especially in the literature that makes contact with AdS/CFT \cite{Maldacena:2002vr,Maldacena:2011nz}, we remind the reader that it is not normalizable when the norm is given by \eqref{finalfixedexpecA}. 
\end{enumerate}

\subsubsection{Dependence on the gauge choice }

The prescription \eqref{ccprescription} defines the expectation value of a {\em gauge-fixed} operator. Since the product of operators $\cprod^{p,q}_{\vI\vJ}$  is not diff$\times$Weyl invariant, if one were to choose a different gauge (as opposed to the transverse-traceless gauge chosen above), one would obtain a different answer for the cosmological correlator. In fact, it is perfectly reasonable to make a different gauge choice, and alternative gauges are discussed in Appendix \ref{app:residual}.  The transverse-traceless gauge is convenient for us since it will be shown below that the symmetries of cosmological correlators take on a simple form. In other gauges, these symmetries might be realized nonlinearly although different gauges might be suitable for different physical applications.

Given a gauge choice, the prescription \eqref{ccprescription} defines an unambiguous conjugate-bilinear functional on two states. Therefore, there necessarily exists {\em some} gauge invariant operator on the Hilbert space whose matrix elements are defined by \eqref{ccprescription}. More precisely, with $|\Psi \rangle = a |\Psi_1 \rangle + b |\Psi_2 \rangle$, we can simply define a gauge-invariant operator $\hat{\cprod}^{p,q}_{\vI\vJ\vec{x}} $ with matrix elements as follows
\be
\label{gaugeinvop}
( \Psi_1 , \hat{\cprod}^{p,q}_{\vI\vJ\vec{x}} \,\Psi_2 ) \equiv  {\partial \over \partial a^*} {\partial \over \partial b} \llangle \Psi | \cprod^{p,q}_{\vI\vJ}(\vec{x}) | \Psi \rrangle_{\text{CC}}~,
\ee
where the right hand side is defined by \eqref{ccprescription}. The operator $\hat{\cprod}^{p,q}_{\vI\vJ\vec{x}}$ is not a local functional of $\chi$ and $h_{i j}$ and the $\vec{x}$ are simply labels for this operator.

The map between the product $\cprod_{p,q}(\vec{x})$ and the gauge invariant operator depends on the gauge choice. However, the difference between different gauge choices manifests itself only at $\Or[\kappa]$. In the nongravitational limit, there is a simple gauge-invariant operator whose expectation value yields \eqref{ccprescription}.  This is given by simply taking the group average of \eqref{ccprescription}. 

To see this more precisely, let $U$ be the operator in nongravitational quantum-field theory that implements the action of the conformal group on the late-time metric and matter fluctuations. Then we have
\be
\label{nggaugeinvop}
\hat{\cprod}^{p,q}_{\vI\vJ \vec{x}} =  {1 \over \text{vol}(\sodminone)} \int dU\,U^{\dagger} \cprod^{p,q}_{\vI\vJ}(\vec{x}) U, \qquad (\kappa \to 0)~,
\ee
where $dU$ is the associated Haar measure. The right hand side makes sense provided $p + q \geq 3$. We see that $\hat{\cprod}^{p,q}_{\vI\vJ \vec{x}}$ is an average of an infinitely delocalized operator. 

In the nongravitational limit, it may be checked using \eqref{higuchingnorm} that the expectation value of \eqref{nggaugeinvop} in a state of the form \eqref{freestates} is the same as \eqref{ccprescription}. We find that
\be
\begin{split}
( \Psi_{\text{ng}} , \hat{\cprod}^{p,q}_{\vI\vJ \vec{x}} \,\Psi_{\text{ng}} ) &= {1 \over \text{vol}(\confgrp)} \int d U \langle 0 | \dcoeff[n,m]  U^{\dagger} \cprod^{p,q}_{\vI, \vJ}(\vec{x}) U \dcoeff[n,m]^*|0 \rangle_{\text{QFT}}  \\
&= \langle 0 | \dcoeff[n,m] \cprod^{p,q}_{\vI,\vJ}(\vec{x}) \dcoeff[n,m]^* | 0 \rangle_{\text{QFT}} \\
&= \ln \Psi_{\text{ng}} |\cprod^{p,q}_{\vI,\vJ}(\vec{x}) |\Psi_{\text{ng}} \rn_\r{QFT}~,
\end{split}
\ee
where, in the second line, we use the invariance of $|\Psi_{\text{ng}} \rangle$ under conformal transformations.

At nonzero $\kappa$ we do not know of any simple analogue of \eqref{nggaugeinvop} that gives an explicit expression for the gauge-invariant operator whose matrix elements coincide with the gauge-fixed operator. Note also that,  at nonzero $\kappa$,  one must take  a linear combination of an infinite set of terms \eqref{cprod} with increasing values of $p$ to construct a gauge-invariant operator of the form given in \eqref{coeffdef}.

\subsection{Symmetries of cosmological correlators \label{subsecsymm}}
Cosmological correlators are defined in \eqref{ccprescription} by inserting a product of operators in the path integral weighted with a specific action.  This action is invariant under the residual gauge transformation that are left unfixed in \eqref{ccprescription}. We will utilize the finite action of translations, rotations and dilatations on the matter fields and the ghosts, which  is given by
\be
\label{ccsymmetries}
\begin{split}
\text{translations:}&\quad  h_{i j}(x) \rightarrow h_{i j}(x + \transv); \quad \chi(x) \rightarrow \chi(x+\transv); \\ &\quad c^i(x) \rightarrow c^i(x+\transv); \quad \bar{c}^i(x) \rightarrow \bar{c}^i(x+\transv); \\ \\
\text{rotations:}&\quad h_{i j}(x) \rightarrow R^{k}_{i} R^{\l}_{j} h_{k\l}(R \cdot x); \quad \chi(x) \rightarrow \chi(R \cdot x);\\ &\quad  c^i(x) \rightarrow R^{i}_{j} c^j(R \cdot x); \quad \bar{c}^i(x) \rightarrow R^{i}_{j} \bar{c}^j(R \cdot x); \\ \\
\text{dilatations:} &\quad h_{i j}(x) \rightarrow  h_{i j}(\lambda x); \quad \chi(x) \rightarrow \lambda^{\Delta} \chi(\lambda x); \\ &\quad c^{i}(x) \rightarrow \lambda^{-1} c^i(\lambda x); \quad \bar{c}^{i}(x) \rightarrow \lambda^{d-1} \bar{c}^i(\lambda x) ~.
\end{split}
\ee
Here $\transv$ is a vector, $R$ is a SO(d) rotation matrix and $\lambda$ is a number.  We discuss special conformal transformations below. It is important that what we call ``dilatations'' above includes not just a diffeomorphism but a compensating Weyl transformation that preserves the gauge conditions. For this reason, the metric transforms as $g_{i j}(x) \rightarrow g_{i j}(\lambda x)$ and does not pick up an overall factor of $\lambda^{-2}$. The metric fluctuation transforms as shown above.

Since the ``action'' $\cS$ is invariant under the transformations above, cosmological correlators transform covariantly under these transformations. Under a combined dilatation and translation we find that in any physical state $|\Psi \rangle$
\be
\label{csymshiftscal}
\llangle \Psi | \cprod^{p,q}_{\vI\vJ}(\lambda \vec{x} + \transv) | \Psi \rrangle_{\text{CC}} = \lambda^{-q \Delta} \llangle \Psi | \cprod^{p,q}_{\vI\vJ}(\vec{x}) | \Psi \rrangle_{\text{CC}} ~.
\ee
Under a rotation we find that
\be
\label{csymrot}
\llangle \Psi | \cprod^{p,q}_{\vI\vJ}(R \cdot \vec{x}) | \Psi \rrangle_{\text{CC}} = R^{i'_1}_{i_1} R^{j'_1}_{j_1} \ldots R^{i'_p}_{i_p}R^{j'_p}_{j_p}  \llangle \Psi | \cprod^{p,q}_{\vI'\vJ'}(\vec{x}) | \Psi \rrangle_{\text{CC}}~.
\ee

The symmetries of cosmological correlators should be distinguished from the symmetries of the coefficient functions \eqref{coeffdef} that appear in the wavefunctional. Those coefficient functions are constrained by the full conformal group, even away from $\kappa \rightarrow 0$, as a consequence of the WDW equation. Cosmological correlators are obtained by squaring and integrating the wavefunctional with a choice of gauge.

\paragraph{\bf Dilatations. }
The inclusion of dilatations in the group of symmetries requires explanation since, in a quantum field theory, scale invariance is often broken by loop effects. So the reader might worry that UV effects might force us to use a regulator that is inconsistent with scale invariance.

However, here, the residual group of symmetries involving dilatations is a subgroup of the \diffWeyl{}  group. The latter symmetry is a gauge symmetry of the path integral used to compute expectation values. So we expect that even if counterterms need to be added to the expression for the wavefunctional to regulate UV divergences, $|\Psi[g,\chi]|^2$ will still remain \diffWeyl{} invariant.  Moreover, the form of the ghost action is protected by BRST symmetry.  Therefore we expect that the reduced Faddeev-Popov determinant that appears in \eqref{ccprescription} remains invariant under these symmetries even when loop effects are included.

\paragraph{\bf Special conformal transformations.}
For $d > 2$ and away from $\kappa \rightarrow 0$ the action of special conformal transformations is corrected as shown in \eqref{residualgrav}. Such a transformation acts on an insertion in \eqref{ccprescription} via\be
\label{actionsct}
\delta_{\xi} \chi = \xi^{i} \partial_i \chi + \frac{\Delta}{d} \nabla_i\xi^i\chi; \qquad \delta_{\xi} g_{i j} = (P \xi)_{i j}~,
\ee
where $P$ is defined in \eqref{pdef}.

But since $\xi^i$ contains factors of the metric, this transformation acts nonlinearly on the fields. In Appendix \ref{app:residual}, it is shown how $\xi^i$ corresponding to SCTs can be found perturbatively in terms of the metric fluctuation. Keeping this structure in mind, we see that the action of \eqref{actionsct} converts a single insertion of the metric or a matter field to an infinite series that involves powers of the metric.  Therefore special conformal transformations relate low-point cosmological correlators to higher-point cosmological correlators \cite{Hinterbichler:2013dpa,Kundu:2015xta}. Although this is an important and useful constraint on cosmological correlators, it will not be required for our purposes.

In the nongravitational limit and in $d=2$, the metric-dependent term in SCTs goes away. So, in that setting, cosmological correlators with a fixed value of $p,q$ transform covariantly under SCTs.

We note that rotations, dilatations and translations act in a simple manner on cosmological correlators because they correspond to metric-independent residual gauge-transformations left unfixed by the transverse-traceless gauge. In some choices of gauge, such as the alternative gauge discussed in Appendix \ref{app:residual}, all the residual gauge transformations are metric dependent. In such a gauge, all the symmetry transformations of cosmological correlators will change the value of $p$. Physically, cosmological correlators are still constrained by these symmetries in such gauges. But the constraints are more complicated than \eqref{csymrot} and \eqref{csymshiftscal}.

\subsection{Symmetries and initial conditions}
Our analysis of symmetries does {\em not} assume that the state in \eqref{ccprescription} is the Euclidean vacuum, as obtained from the Hartle-Hawking proposal. Cosmological correlators in all states have the same symmetries; and vacuum cosmological correlators do not display an enhanced symmetry group. 

This also means that, contrary to what is sometimes claimed, the observed approximate symmetries of cosmological correlators including scale invariance do {\em not} provide evidence that our universe was in the Hartle-Hawking state during the inflationary period.  If there was a period of inflation, and if the universe was well described by an excited state of the form \eqref{psiderstate}, one would obtain cosmological correlators with the same symmetries. 

This strengthens the argument made in \cite{Mata:2012bx} that the symmetries of correlators --- which completely fix specific low-point functions --- provide a sharp test of inflation, since it removes the need for assuming a particular initial state.

On the other hand, to make contact with empirical observations, it is often interesting to consider departures from the slow-roll approximation. These must be present in the real world since inflation cannot go on forever but must end before the local curvature becomes arbitrary small. To analyze these corrections in our language, would require knowledge of the state away from the large-volume limit.  We are unable to make any statements about these corrections since we have not considered these subleading terms in this paper or in  \cite{dsfirst2023}.

\section{Holography of information \label{secholinfo}}

The symmetries of cosmological correlators immediately lead to a remarkable result. 

Let $\reg$ be any open subset of $\R^d$ and $\vec{x}'= (x_1',\dots,x_{p+q}')$ be an arbitrary set of $p + q$ points in $\R^d$. Then we can find a set $\vec{x}$  of $p + q$ points in $\cR$ such that
\be
x_k'  = \lambda x_k + \transv,\qq x_i\in \cR,\qq k=1,\dots,p+q~,
\ee
for some choice of $\la>0$ and vector $\transv$. In other words, an arbitrary configuration of points can always be mapped to lie in the region $\cR$ with a suitable dilatation and translation.

Therefore, if we are given the set of all cosmological correlators
\be
\{\llangle \Psi | \cprod^{p,q}_{\vI\vJ}(\vec{x}) |\Psi \rrangle_{\text{CC}}\}
\ee
for all values of $p,q$ and all configurations of points $x_i \in \reg$, the symmetry \eqref{csymshiftscal} implies that this information is sufficient to determine all cosmological correlators in the state $\Psi$ everywhere on the spatial slice.  But the set of all cosmological correlators everywhere on the spatial slice are evidently enough to reconstruct all observables on the slice.

This immediately leads us to the following result.
\begin{result}
The set of all cosmological correlators in any open region $\reg$ in a state $\Psi$  determines all observables in the state.
\end{result}

Our result relies on the relation \eqref{csymshiftscal}. In other gauge choices, such as the alternative gauge discussed in Appendix \ref{app:altgauge}, translations and dilatations will act on cosmological correlators by changing the value of $p$ as the residual symmetry generators have metric-dependent corrections. Nevertheless, they still relate the set of all cosmological correlators in a region $\reg$ (\ie cosmological correlators with all possible value of $p,q$) to cosmological correlators outside that region. Therefore we expect that the result above should also hold for cosmological correlators in such gauges although it is calculationally harder to obtain the value of a cosmological correlator outside $\reg$ using information inside $\reg$.

\subsection{Nongravitational limit} 
Somewhat surprisingly, the result above remains true even as we take $\kappa \rightarrow 0$. 

It was shown in \cite{dsfirst2023} that the states $\Psi_{\text{ng}}$ (displayed in \eqref{freestates}) have the property that they are invariant under the de Sitter isometries,
\be
U |\Psi_{\text{ng}} \rangle = |\Psi_{\text{ng}} \rangle~.
\ee
Following the steps in subsection \ref{subsecnormnongrav}, we see that the expression for the cosmological correlator is simply
\be
\lim_{\kappa \to 0} \,\llangle \Psi_{\text{ng}} | \cprod^{p,q}_{\vI\vJ}(\vec{x}) | \Psi_{\text{ng}} \rrangle_{\text{CC}} =    \langle \Psi_{\text{ng}} | \cprod^{p,q}_{\vI\vJ}(\vec{x}) |\Psi_{\text{ng}} \rangle_{\text{QFT}}~,
\ee
where, on the right hand side, we now find simply the QFT expectation value of $\cprod^{p,q}_{\vec{i} \vec{j}}(\vec{x})$ in the state $|\Psi_{\text{ng}} \rangle$. Using the invariance of the state under the de Sitter isometries we see that
\be
\llangle \Psi_{\text{ng}} | \cprod^{p,q}_{\vI\vJ}(\vec{x}) | \Psi_{\text{ng}} \rrangle_{\text{CC}} = \llangle \Psi_{\text{ng}} | U^{\dagger} \cprod^{p,q}_{\vI\vJ}(\vec{x}) U | \Psi_{\text{ng}} \rrangle_{\text{CC}}~.
\ee
So, in the nongravitational limit cosmological correlators are invariant under the entire conformal group. This includes the action of special conformal transformations that do not appear in the group of symmetries at finite $\kappa$ shown in \eqref{ccsymmetries}. The result on the holography of information follows immediately.

Physically, this analysis tells us that  holography of information does not rely on the measurement of ``small gravitational tails''  but rather from an imposition of the gravitational Gauss law. The constraints implied by the Gauss law restrict the form of the allowed states in the theory, which is why it is possible to uniquely identify states from cosmological correlators in any open set.

\subsection{Difference between quantum field theories and quantum gravity}
We have shown that holography of information persists, if one takes the nongravitational limit of a theory of gravity while preserving the gravitational Gauss law.  We now explain why nongravitational quantum field theories do not display this property. 

Starting with the Euclidean vacuum, which is still obtained by the Hartle-Hawking prescription, states in a QFT take the form
\be
\label{qftstates}
|\qstate  \rangle = \int d \vec{y}d\vec{z}\, \qstate^{\vI\vJ}(\vec{y}, \vec{z})  h_{i_1 j_1}(y_1) \ldots h_{i_n j_n}(y_n) \chi(z_1) \ldots \chi(z_m) |0 \rangle~,
\ee
where $h_{i j}$ are transverse traceless graviton fluctuations. Here $\qstate^{\vI\vJ}$ is an arbitrary smearing function and the only constraint is that $|\qstate \rangle$ should be normalizable under the usual QFT norm.
\be
\label{qftnorm}
\begin{split}
\langle \qstate |\qstate\rangle_{\text{QFT}} = \int &d \vec{x} d \vec{x}'  \qstate^{\vI\vJ}(\vec{y}, \vec{z})^* \qstate^{\vI'\vJ'}(\vec{y}', \vec{z}') \\ &\times \langle 0| h_{i'_1 j'_1}(y'_1)\ldots \chi(z'_m) h_{i_1 j_1}(y_1) \ldots \chi(z_m) |0 \rangle_{\text{QFT}} ~.
\end{split}
\ee
  We emphasize the difference with the Hilbert space obtained in the nongravitational limit of a gravitational theory where the states take the form \eqref{freestates}.  In \eqref{freestates} the smearing function is constrained by conformal symmetry, whereas in \eqref{qftstates} it is not.

Moreover, the smearing functions that appear in \eqref{freestates} are {\em disallowed} by normalizability in \eqref{qftstates}.  This is simply the statement that, apart from the vacuum, there are no states that are invariant under the de Sitter isometry group in the usual QFT Hilbert space. This can also be seen directly from the expression for the norm \eqref{qftnorm}. The correlator in the Euclidean vacuum is conformally covariant because the Euclidean vacuum itself is invariant. But if this correlator were to be integrated with the smearing function that appears in \eqref{freestates} the entire integrand would be invariant under the action of the conformal group. Therefore, the norm would pick up a divergence proportional to the volume of the conformal group. When we consider states in the nongravitational limit of a gravitational theory and use the correct norm, this divergence is cancelled by dividing by the volume of the conformal group  but there is no such factor in the ordinary QFT norm.

Therefore for a generic value of $\lambda$ and $\transv$ and for any QFT state except for the vacuum,
\be
\langle \qstate| \cprod^{p,q}_{\vI\vJ} (\lambda \vec{x} + \transv)  |\qstate \rangle_{\text{QFT}} \neq  \lambda^{-q \Delta } \langle \qstate| \cprod^{p,q}_{\vI\vJ}(\vec{x})  |\qstate \rangle_{\text{QFT}}~.
\ee
So the argument leading to the holography of information breaks down in the QFT Hilbert space.

As usual, in a QFT, it is possible to prepare ``split states''\cite{Haag:1992hx} where correlators coincide inside a region but differ outside that region. This means the following. Let $\overline{\reg}_{\epsilon}=\overline{\reg \cup \epsilon}$ be the complement of the union of the region $\reg$ and a small ``collar region'', $\epsilon$. Then given any two states of the form \eqref{qftstates} one can find a split state with the property that when $x_i \in \reg$ and $x'_i \in \overline{\reg}_{\epsilon}$
\be
\langle \qstate^{\text{split}} | \cprod^{p,q}_{\vI\vJ}(\vec{x}) \cprod^{p',q'}_{\vI'\vJ'}(\vec{x}') | \qstate^{\text{split}} \rangle_{\text{QFT}} = \langle \qstate_{1}| \cprod^{p,q}_{\vI\vJ}(\vec{x}) | \qstate_{1} \rangle_{\text{QFT}} \langle \qstate_{2} | \cprod^{p',q'}_{\vI'\vJ'}(\vec{x}') | \qstate_{2}\rangle_{\text{QFT}} ~
\ee
for any choice of the operators $\cprod^{p,q}_{\vI\vJ}(\vec{x})$ and $\cprod^{n',m'}_{\vI'\vJ'}(\vec{x}')$. 
In such a split state, not only are observations in $\overline{\reg}_{\epsilon}$ not determined by observations in $\reg$, they are not even correlated. Clearly this means that the full state cannot be identified by observations in $\reg$.

We conclude that the result on holography of information marks a clear mathematical difference between the properties of quantum field theory and quantum gravity, in terms of how such theories localize information. This difference persists in the nongravitational limit of a gravitational theory provided one consistently imposes the Gauss law while taking this limit.

\subsection{Comparison to flat space and AdS}
The result above can be placed in the context of similar results proved in AdS and in asymptotically flat space. There, the principle of holography of information is usually framed as follows: ``the information in the bulk of a Cauchy slice is also available near its boundary.'' More precisely, in asymptotically flat space, it was shown in \cite{Laddha:2020kvp} that all information that is available on all of ${\scrip}$ is also available on its past boundary $\scrippast$;  and, in a spacetime that is asymptotically AdS, all information that is available on the timelike boundary is also available in an infinitesimal time band. 

In the form above, it is unclear how the principle should be generalized to dS, where a Cauchy slice has no boundary. But, consider the following alternative phrasing of this principle: ``in all pure states of the theory, whenever a region, $\reg$,  is completely surrounded by its complement, $\overline{\reg}$, then all the information inside $\reg$ is accessible in $\overline{\reg}$.''\footnote{We restrict to pure states to avoid situations where entanglement with an auxiliary system has produced an ``island'' inside $\reg$.} In flat space and AdS, this is trivially equivalent to the usual statement; when $\overline{\reg}$ surrounds $\reg$ then it also includes the asymptotic region near infinity.

\begin{figure}[!h]
\begin{center}
	\begin{tabular}{cc}  
		\subf{\includegraphics[height=4.4cm]{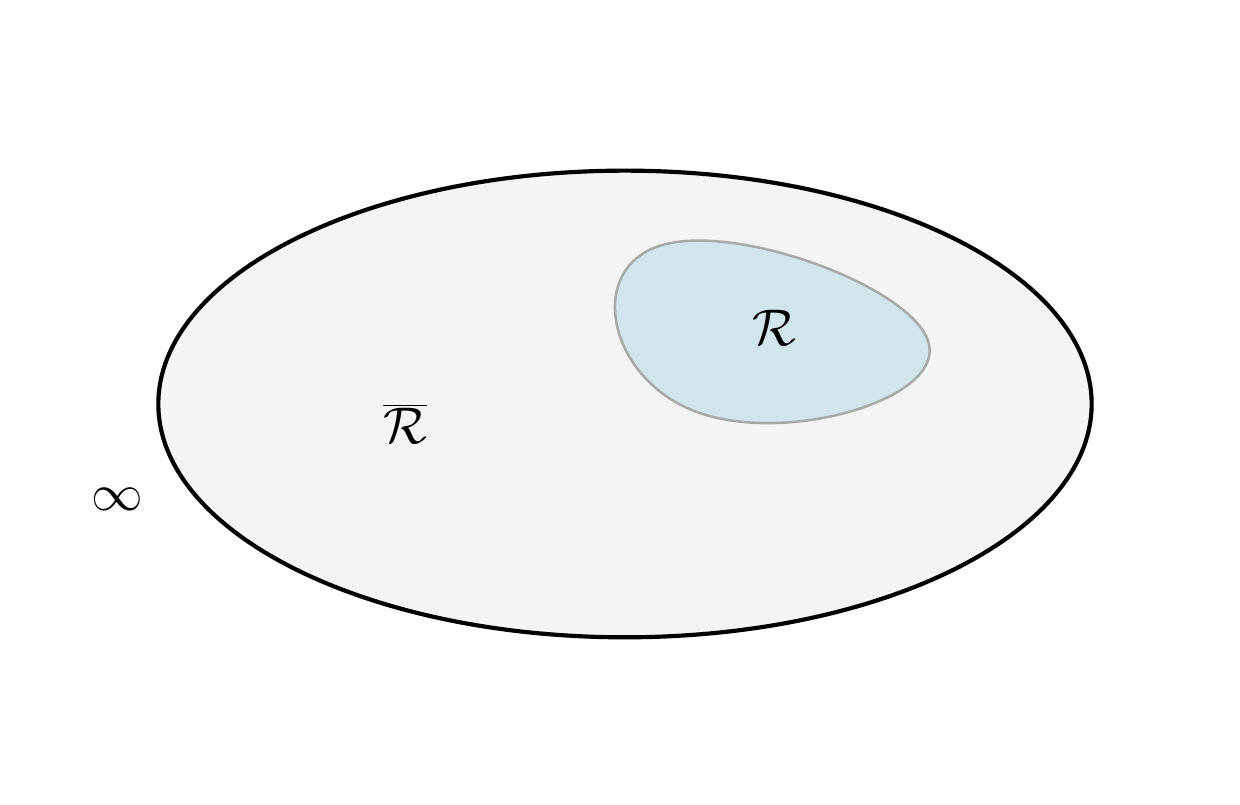}}{AdS or flat space} \hspace{1cm} &
		\subf{\includegraphics[height=4.4cm]{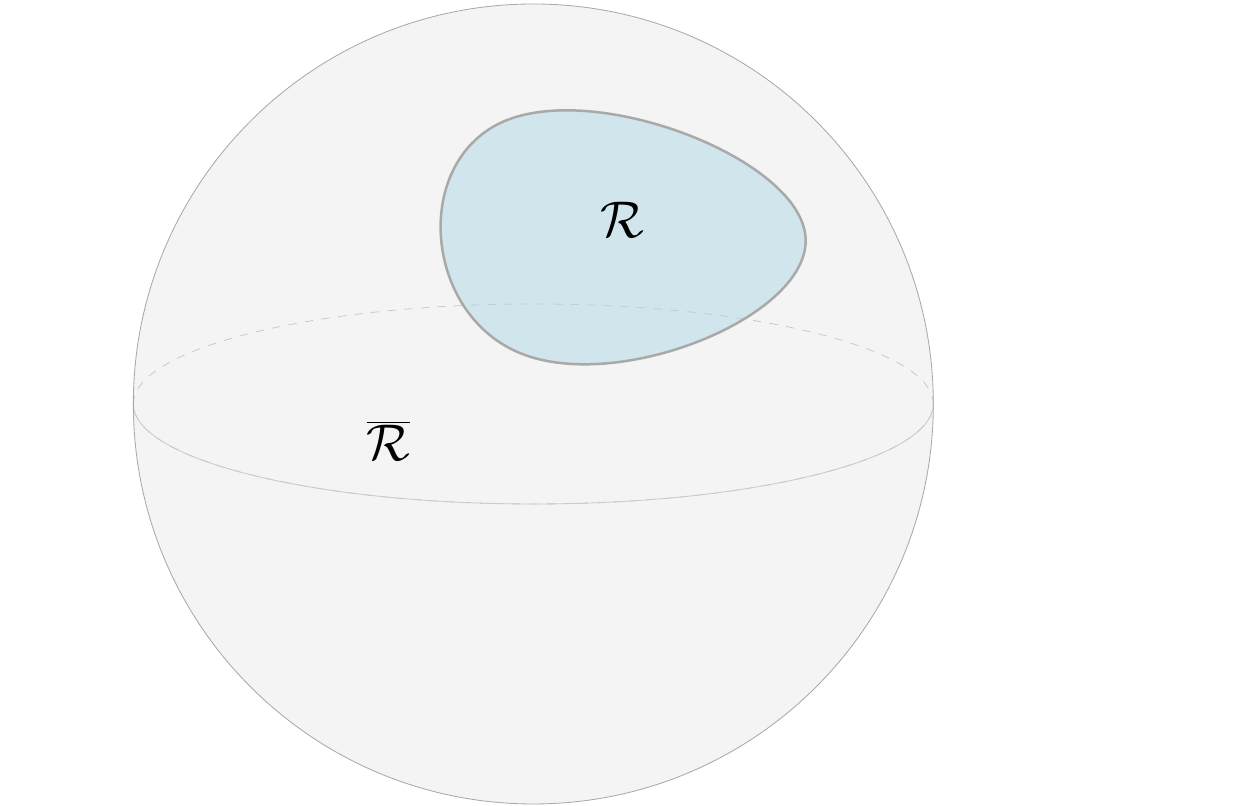}}{dS} 
	\end{tabular}
	\caption{\em In flat space and in AdS (left), when a region on a spatial slice, $\reg$ is surrounded by its complement then $\overline{\reg}$ extends to infinity. But in dS (right), every region $\reg$ surrounds and is surrounded by its complement on a sphere. \label{figdsreg}}
\end{center}
\end{figure}

The second form of the slogan generalizes naturally to dS.  Since the Cauchy slices in dS are compact, every region $\reg$ both surrounds its complement and is surrounded by its complement. (See Figure \ref{figdsreg}.) So it is natural that cosmological correlators in  every region $\reg$ contain all the information that is available on the Cauchy slice in a pure state.

\subsection{Higher-spin matter fields and stringy corrections}

In the analysis above, we have studied a massive scalar field in the matter sector. This choice was made for simplicity. It seems clear that the proof of the principle of holography of information will go through in the presence of higher-spin matter. 

Our results in \cite{dsfirst2023} and in this paper rely only on an {\em asymptotic analysis}. The assumption is that the formalism of quantum field theory makes sense at asymptotic infinity. This assumption is usually taken to be valid even in the presence of stringy corrections.\footnote{Here, we do not enter into the recent debates on whether de Sitter solutions can be found within string theory \cite{Kachru:2003aw,Garg:2018reu,Obied:2018sgi}.}

However, there is an important difference in dS compared to AdS and flat space. In the latter setting, it is reasonable to assume that the asymptotic structure of the spacetime is not modified even nonperturbatively. Therefore the results of \cite{Laddha:2020kvp} are expected to hold even nonperturbatively. But the asymptotic structure of dS is not expected to be nonperturbatively stable \cite{Goheer:2002vf}. Therefore, nonperturbatively, our analysis might require modifications.

\subsection{Cautionary physical remarks}

The principle of holography of information provides an interesting mathematical difference between quantum field theories and quantum gravity, but the result should be interpreted with care. First, as we have emphasized above, there are no {\em local} gauge invariant operators in the theory. Therefore, the measurement of a cosmological correlator is secretly a nonlocal process. Cosmological correlators are labelled by a set of points in $\reg$; but they do not correspond to any physical observable that is strictly localized in $\reg$.

Second, in both AdS and flat space, if one considers heavy, nonperturbative states in the bulk, then it is usually necessary to study nonperturbative correlators at infinity to identify the state. This point was already noted in \cite{Laddha:2020kvp,Chowdhury:2021nxw} and recently re-emphasized in \cite{Bahiru:2023zlc,Bahiru:2022oas}. So, in a typical heavy classical state, mundane notions of locality are preserved at all orders in perturbation theory. This is important since it explains why we do not ``see'' the holography of information all around us.

This does not mean that the unusual localization of information in gravity is unimportant. In its nonperturbative avatar, it is important for understanding the information paradox \cite{Raju:2020smc}.   Moreover, if one studies simple states like low-energy excitations about empty AdS then the holography of information can be seen even within perturbation theory. 

We expect the same features to hold in dS. In a ``little Hilbert space'' comprising simple excitations about the Hartle-Hawking state, we expect it should be possible to identify states uniquely using only perturbative cosmological correlators. On the other hand, to identify sufficiently complicated states might require very high-point cosmological correlators. It would be interesting to work this out in more detail.

\section{Discussion \label{secdiscussion}}
In this paper, we started by studying the norm on the space of solutions to the WDW equation obtained in \cite{dsfirst2023}. The magnitude-squared of these wavefunctionals leads to a \diffWeyl{} invariant functional. We defined the norm by averaging this functional over field configurations and dividing  by the volume of the \diffWeyl{} group.  We used the Faddeev-Popov trick to make sense of this expression, leading to the final gauge-fixed expression \eqref{finalfixedexpecA}.

In the nongravitational limit, our norm reduces to the one proposed by Higuchi on the space of group-averaged states. Therefore,  our procedure provides a derivation of Higuchi's prescription in the nongravitational limit and a means of understanding gravitational corrections to this prescription.

In section \ref{secholinfo}, we explored the meaning of cosmological correlators. We proposed that these commonly-discussed quantities correspond to gauge-fixed observables. These observables are labelled by a set of local coordinates although their gauge-invariant description is necessarily nonlocal. We showed that, in any state of the theory, these observables are invariant under rotations, translations and dilatations of their coordinate labels.  This marks a sharp difference from nongravitational quantum field theories, where cosmological correlators manifest this symmetry in the vacuum but not in other states.  As a consequence of this symmetry, we showed that, in a theory of gravity, cosmological correlators in an arbitrarily small region, $\reg$, suffice to uniquely identify any state in the theory.

These results open up several interesting questions that we now describe.

\paragraph{\bf Holography in de Sitter.} Strictly speaking, our result on the holography of information does not allow us to obtain information about a higher-dimensional space from a lower-dimensional space since $\reg$ still has codimension $0$.  This is similar to the situation in AdS --- where arguments based on the gravitational constraints are sufficient to show that information in the bulk is available in an infinitesimal time band in the boundary, but are not sufficient to squeeze the time band to a time slice. Moreover, our results pertain to information but do not address the issue of bulk dynamics.

So the natural question is whether there is some way of understanding bulk dynamics in all of de Sitter space from a lower-dimensional subregion on the late-time slice. Similar ideas were recently explored in \cite{Moitra:2022glw}. (See \cite{Pasquarella:2022ibb} for a different perspective.)

 Such a holographic duality, if it exists, should account for all states in the bulk theory. In the literature, the study of dS/CFT has often been restricted to understanding the Euclidean vacuum, obtained from the Hartle-Hawking proposal. But as we have shown the bulk theory has many other interesting states.

It would also be interesting to understand the relationship of such a holographic dual to the work on holographic spacetime \cite{Banks:2001yp,Banks:2018ypk,Banks:2020zcr} and the proposal of static patch holography \cite{Banks:2006rx,Susskind:2021dfc}. In the latter proposal, it is suggested, using very different arguments,   that all information about the state can be obtained from the bifurcation sphere that lies between two static patches. This sphere lies in the ``bulk'' of dS whereas our results have to do with the asymptotic late-time slice. So our results do not contradict this proposal, but nor do they obviously lend it support.

\paragraph{\bf Observers in quantum cosmology. }
An interesting conceptual question is the following. Gauge-invariant operators in gravity must be nonlocal but this is in apparent contradiction with our physical intuition that measurements are made locally. Fixing the gauge, as we did to study cosmological correlators,  provides a mathematically convenient method of obtaining observables that are labelled by a set of coordinates. But it is important to develop a deeper understanding of the meaning of measurements in a cosmological setting. The usual theory of measurement \cite{preskill1998lecture} involves an external apparatus that is entangled with the system by the experimenter who turns on an interaction Hamiltonian. Clearly this cannot correctly describe measurements in a theory of gravity, where bulk evolution is generated by the constraints, that cannot be altered at will. Presumably, the correct framework is to study an observer who is already part of the system and where measurement happens through the {\em autonomous evolution} of the system. We do not know the correct formalism to analyze this process.

A simple model of an observer was recently discussed in \cite{Chandrasekaran:2022cip} where it was argued that the algebra of observables dressed to the observer's worldline is of type II$_1$. Since we have presented the full Hilbert space and a formalism for understanding observables, it should be possible to embed the model of \cite{Chandrasekaran:2022cip} into our analysis and make it precise. It would be interesting to work out these details.

\paragraph{\bf Technical questions about the norm. }
From a technical perspective, we would like to better understand the functional integral that was used to define the norm. Some subtleties, including the question of the measure, the requirement of a minimum of three operator insertions,  and potential divergences due to the ``collision'' of operators  are listed in subsection \ref{secsubtlenorm}. Similar problems have been studied extensively in string perturbation theory and we hope that the techniques developed there can be applied to the functional integrals that appear in our context.

\paragraph{\bf Implications for cosmology.}
 Our result implies that when gravitational constraints are taken into account, every physical state has the same symmetries as the vacuum. This would not be true in quantum field theory where the vacuum is singled out by its symmetries. This suggests that the approximate scale invariance observed in the early universe cannot be used to justify the Hartle-Hawking proposal. It is in fact a general consequence of the constraints in any asymptotically de Sitter spacetime, such as the early universe as predicted by inflation.

\section*{Acknowledgments}
We are grateful to Simon Caron-Huot, Abhijit Gadde, Rifath Khan, Alex Maloney, Ashoke Sen and Sandip Trivedi for helpful discussions. We also acknowledge several discussions with the string theory group at ICTS-TIFR. S.R. would like to acknowledge the hospitality of the 12th Joburg Workshop on string theory, the Abu Dhabi meeting in theoretical physics and the workshop on observables in quantum gravity (IISER Mohali) where preliminary versions of these results were presented. S.R. is partially supported by a Swarnajayanti fellowship, DST/SJF/PSA-02/2016-17, of the Department of Science and Technology. J.C. is supported by the Simons Collaboration on Nonperturbative Bootstrap. Research at ICTS-TIFR is supported by the Department of Atomic Energy, Government of India, under Project Identification Nos. RTI4001.

\section*{Appendix}
\appendix
\changelocaltocdepth{1}

\section{Residual gauge symmetry}\label{app:residual}

In this Appendix, we study the residual gauge symmetry after fixing the diffeomorphism and Weyl gauge symmetries using
\begin{equation}
\label{gaugechoiceapp}
	\partial_i g_{ij}=0, \qq\delta^{ij}g_{ij}= d~.
\end{equation}

After solving for the traceless condition, the variation of the metric is a combination of a diffeomorphism and a Weyl transformation 
\begin{equation}
	\delta_{\xi}g_{ij } = (P\xi)_{ij} \equiv \cL_\xi g_{ij} -{1\/d} g_{ij}\d^{k\l}\cL_\xi g_{k\l}
\end{equation}
in terms of the Lie derivative
\be
\cL_\xi g_{ij} = \xi^k\p_k g_{ij}+ g_{ik}\p_j\xi^k+g_{kj}\p_i\xi^k = \n_i\xi_j+\n_j\xi_i~.
\ee
The residual gauge symmetry algebra corresponds to solutions of 
\be\label{appresconditionxi}
\p_i (P\xi)_{ij}= 0~.
\ee
The metric is written as
\be
g_{ij} = \d_{ij}+\k h_{ij}~,
\ee
which leads to the expansion
\be
(P\xi)_{ij} = (P_0\xi)_{ij}+ \k (P_1\xi)_{ij}+\k^2 (P_2\xi)_{ij}~,
\ee
that is exact since no higher orders of $\kappa$ appear.

Firstly, note that in the limit $\k\to 0 $, the residual symmetry is $\r{SO}(1,d+1)$ because we then have
\be
(P_0\z)_{ij} = \p_i \z_j+ \p_j\z_i - {2\/d}\d_{ij}\d^{k\l}\p_k\z_\l =0~,
\ee
for any conformal Killing vector $\z$. In other words, conformal Killing vectors preserve the background  metric and hence trivially preserve any gauge-fixing condition.

\subsection{Translations, rotations and dilatations}

We will see that translations, rotations and dilatations remain residual symmetries at finite $\k$. To show this, we write the explicit form 
\be
\label{pipexplicit}
\p_i (P\xi)_{ij} = \le[\le( \partial_k g_{ij }+ \partial_ j g_{ik }  - \frac{2}{d}g_{i\l }\partial_\l g_{jk }\ri)\p_k + \le( g_{ij } \d_{k\l} + g_{jk }\d_{i\l}  - \frac{2}{d}g_{ik } g_{j\l} \ri)\partial_k \partial_\l\ri]\xi^j ~.
\ee
We see that translations $\xi^j = \r{const}$ are always a residual symmetry since at least one derivative acts on $\xi^j$ in \eqref{pipexplicit}.   For rotations and dilatations, the term with two derivatives vanishes so we get
\be
\p_i (P\xi)_{ij} = \le( \partial_k g_{ij }+ \partial_ j g_{ik }  - \frac{2}{d}g_{i\l }\partial_\l g_{jk }\ri) \partial_k \xi^j ~.
\ee
Rotations are of the form $\xi^j = M^{jk} x^k$  where $M^{jk}$ is antisymmetric. So we see that $\partial_k \xi^j = M^{jk}$ and the above expression vanishes by symmetry.  For dilatations, we have $\xi^j = x^j$ so $\partial_k \xi^j  = \d^{j}_{k}$ and the expression vanishes using the transverse and trace conditions.

As a result, we see that translations, dilatations and rotations are residual symmetries.

\subsection{Modified special conformal transformation}

The usual special conformal transformation takes the form
\be
v_0^i =  2(\b\cdot x)x^i - x^2\b^i ~.
\ee
We can check that this is not a residual symmetry as we have
\be
\p_i (Pv_0)_{ij} = -2\k d \b^j h_{ij}~,
\ee
which does not vanish. 

However, using a standard perturbative procedure, the SCT can be systematically corrected \cite{Ghosh:2014kba} to give a residual symmetry, $\xi$, 
\be
\xi = v_0+  \kappa v_1 + \kappa^2 v_2+\dots~.
\ee

Define the operator
\be
(\cD_0 \xi)^j \equiv \p_i (P_0\xi)_{ij} = \p^2 \xi^j  +\le(1-{2\/d}\ri)\p_i \p_j \xi^i~.
\ee
The corrections $v_n$ are obtained by solving the equation \eqref{appresconditionxi}
\be\label{eqsourceSCT}
(\cD_0 v_n)^j = s_n^j,\qq n=1,2,\dots
\ee
where 
\be
s_1^j = 2 d \b^i h_{ij}~,
\ee
and the higher order sources are determined iteratively using
\be
 s_n^j =- \p_i(P_1 v_{n-1})_{ij} - \p_i(P_2 v_{n-2})~,\qq n\geq 2~.
 \ee

We can show that, provided one places physical boundary conditions on the metric fluctuation, the equation \eqref{eqsourceSCT} always has a smooth solution $v^j$ that is smooth on the sphere so that it is always possible to correct the SCT in this way. As detailed in the next section these boundary conditions constrain the metric around $x=\infty$ to be
\be
h_{ij} = W_{ikj\l}{x_kx_\l\/|x|^4}+ O(|x|^{-3})~,
\ee
where $W_{ijk\l}$ is a constant tensor with the symmetries and tracelessness of a Weyl tensor. As a result the first source has the fall-off
\be
s_1^i =2 d \,\b^i W_{ikj\l}{x_kx_\l\/|x|^4}+ O(|x|^{-3})~,
\ee
and higher order sources are more suppressed as they contain additional factors of the metric.

The decay of the sources at infinity  guarantees that solutions for $v^j$ always exist. For the leading fall-off, we obtain the solution
\be
v^j_1=\b^i W_{ikj\l} { x_k x_\l\/|x|^2}+ O(|x|^{-1}) =|x|^2\b^j h_{ij}+ O(|x|^{-1})~,
\ee
which is just proportional to the source at this order. For the subleading fall-offs, the solution can be written in Fourier space
\be
v^j (x) = \int{d^dp\/(2\pi)^d} \,{1\/p^2}\le( -\d_{ij} +{2(d-2)\/d-1}{ p_ip_j\/p^2} \ri) e^{i px}\hat{s}^i(p)~,
\ee
which is well-defined as the sources are $s^i = O(|x|^{-3})$ at infinity so their Fourier transforms are $\hat{s}^i(p) = O(|p|^{3-d}) $ around $p=0$.

\subsection{Boundary condition for the metric}\label{app:bdycond}

Although the physical metric is the round metric on $S^d$, we have performed a Weyl transformation so that the background metric becomes flat. The Weyl factor is singular at $x=\infty$ so we must impose an appropriate boundary condition at infinity. The physical metric takes the form
\be
ds^2 = {4\/(1+|x|^2)^2}(\d_{ij} + \k h_{ij})d x_i dx_j~
\ee
and we should demand that this metric  be regular. In addition we impose the gauge-fixing conditions
\be
\p_i h_{ij }=0,\qq \d_{ij}h_{ij}=0~.
\ee
The Ricci scalar of the physical metric must be a regular function on the sphere. At first order in $\k$, it is a linear combination of $\p_i \p_j h_{ij}, x_j \p_i h_{ij}$ and $x_i x_j h_{ij}$. The first two terms vanish due to the transverse condition and we obtain
\be
R= {d(d-1)}(1 + \k  h_{ij}x_i x_j)  + O(\k^2)~.
\ee
This must be a smooth function on the sphere which implies that $h_{ij}x_i x_j$ must tend to a constant $C$ at infinity. Our gauge-fixing conditions also imply that
\be
\p_i (x_j h_{ij}) = \d_{ij}h_{ij}+x_j \p_i h_{ij}=0~,
\ee
which after integration over a ball of radius $r$ gives by Stokes' theorem
\be
0 = \int_{B_r} d^d x\, \p_i ( x_j h_{ij}) = \int_{ S_r} d^{d-1}\Om \,r^{d-2} x_i x_j h_{ij} = \r{vol}(S^{d-1}) r^{d-2}C\qq r\to+\infty~.
\ee
This implies that $C=0$ so we find that 
\be\label{limboundarymetric}
\lim_{x\to \infty} h_{ij}x_i x_j =0~.
\ee
Additional constraints come from demanding that the metric be smooth near infinity.

\pg{Expansion around infinity.} The metric around $x=\infty$ can be expanded using the inverted coordinates defined as
\be\label{definverted}
\bx_i = {x_i\/|x|^2}~.
\ee
The inverted metric $\bh_{ij}$ is defined as
\be
ds^2 = {4\/(1+|\inv{x}|^2)^2} (\d_{ij}+\k \inv{h}_{ij})d \inv{x}_i d\inv{x}_j~,
\ee
and is related to the original metric using
\be
h_{ij} = {1\/|x|^4}(\d_{ik}|x|^2 -2 x_i x_k)(\d_{j\l}|x|^2 -2 x_j x_\l) \bh_{k\l}~.
\ee
The expansion around $x=\infty$ is an expansion around $\inv{x}=0$.  The boundary condition \eqref{limboundarymetric} in inverted coordinates gives
\be\label{bcinverted}
\lim_{\bx\to 0} {\bx_i\bx_j  \bh_{ij}\/|\bx|^4}=0~.
\ee
To analyze this condition, we demand that the metric be smooth and at least twice differentiable near $\bx = 0$ so that it is possible to perform a series expansion
\be
\inv{h}_{ij}(\bx) =  H^{(0)}_{ij} + H^{(1)}_{ijk} \bx_k + H_{ij k\l }^{(2)}\inv{x}_k\inv{x}_\l + \ldots~,
\ee
where $H_{ijk\dots}^{(n)}$ are constant tensors. 

For the leading orders, the limit implies that we identically have
\be
H^{(0)}_{ij}\bx_i \bx_j =0,\qq   H^{(1)}_{ijk} \bx_i\bx_j\bx_k =0,\qq  H_{ij k\l }^{(2)}\bx_i\bx_j \inv{x}_k\inv{x}_\l =0~.
\ee
Taking derivatives, we obtain that $H_{ij}^{(0)} = 0$. For the linear term, we obtain constraints on $H^{(1)}_{ijk}$ which allows us to write the transverse equation as
\be
0=\p_i h^{(1)}_{ij} = {(d-1)\/2|x|^4} H_{k\l j}^{(1)}x_k x_\l, 
\ee
which implies that $H^{(1)}_{ijk}=0$. This means that $\lim_{\bx \to 0 }\bd_k \bh_{ij}=0$ so that $\inv{x}_i$ are the Riemann normal coordinates around $\bx=0$. Thus we have
\be
\inv{g}_{ij}=  \inv{g}_{ij}(0)+ {1\/3}\inv{R}_{ikj\l}(0)\bx_k \bx_\l + O(|\inv{x}|^3)~,
\ee
and this fixes the term quadratic  to
\be
H^{(2)}_{ijk\l}={1\/3}\inv{R}_{ikj\l}(0)~.
\ee
The tracelessness condition also implies that $\inv{R}_{ij}(0)=0$ so this is really a Weyl tensor. As a result $H^{(2)}_{ijk\l}$  can be any constant tensor with the same symmetries of a Weyl tensor. Conversely, we can verify that this gives a valid metric.

As a result, we obtain the leading behavior of the metric at infinity
\be
h_{ij}= W_{ikj\l}{x_kx_\l\/|x|^4} + O(|x|^{-3}),\qq x\to+\infty~,
\ee
where $W_{ijk\l}$ is a constant tensor with the symmetries and tracelessness of a Weyl tensor.  Note that for dS$_4$, we have $W_{ijk\l}=0$ as there are no non-trivial Weyl tensor in $d=3$ ; so in this case we have $h_{ij}= O(|x|^{-3})$.

\subsection{Residual symmetry algebra}

The residual symmetry algebra is generated by vector fields $\xi[g]$ which in general have metric-dependent corrections. The Lie bracket between two generators must be modified as
\be\label{modifiedbracket}
[\xi_1[g],\xi_2[g]]_\r{M} = [\xi_1[g],\xi_2[g]] - \d_{\xi_1[g]}\xi_2[g] + \d_{\xi_2[g]}\xi_1[g]~,
\ee
where we have added the action of the transformation on the metric-dependent terms obtained from the transformation of the metric.

For example, let $\xi_1$ be a translation, rotation or dilatation and $\xi_2$ be a modified SCT. We can write
\be
\xi_2 = \z + v[h]~,
\ee 
where $\z$ is the unmodified SCT and $v[h]$ contains the metric-dependent corrections. We then have 
\be
[\xi_1 , \xi_2]_\r{M} = [\xi_1,\z] +[\xi_1,v[h]] - \d_{\xi_1} v[h] = [\xi_1,\z]~,
\ee
which gives the standard Lie bracket as if the SCT was unmodified.

As a result, the modification at finite $\k$ doesn't affect the algebra which is always the conformal algebra. The residual symmetry group is then always $\r{SO}(1,d+1)$. However, the finite $\k$ corrections to the SCT modify the way this group acts on the fields.

\subsection{Alternative gauge-fixing conditions}\label{app:altgauge}

In this paper, we have presented our analysis in a Weyl gauge where the background metric for the sphere is flat. We can also consider the a similar gauge-fixing procedure where we keep the round metric. 

In this case, we write the metric as
\be
g_{ij} = \g_{ij}+ \k h_{ij},\qq \g_{ij} = {4 \d_{ij}\/(1+|x|^2)^2}~,
\ee
where $\g_{ij}$ is the round metric on $S^d$. The gauge fixing conditions can be taken to be
\be
\g^{jk}D_k g_{ij} =0 ,\qq \g^{ij}g_{ij} = d~,
\ee
where we use $D_i$ for the background covariant derivative with respect to $\g_{ij}$.

After solving for the trace condition, the variation of the metric is
\be
\d_\xi g_{ij} = (P\xi)_{ij} = (P_0\xi)_{ij}+\k (P_1\xi)_{ij}+\k^2 (P_2\xi)_{ij}~,
\ee
and the residual symmetry is generated by solutions of
\be
\g^{jk}D_k(P\xi)_{ij} =0 ~.
\ee
Again we see that at $\k\to0$, the residual symmetry is generated by the CKVs as we have
\be
(P_0\xi)_{ij} =D_i \xi_j +D_j\xi_i - {2\/d}\g_{ij}\g^{k\l}D_k\xi_\l
\ee
is the conformal Killing equation on $S^d$.

At finite $\k$, we can write
\be
\xi^i = v_0^i+ \k v_1^i +\k^2 v_2^i+\dots~.
\ee
Taking $v_0$ to be any CKV, we can make $\xi$ into a residual symmetry by choosing the corrections $v_n$ to be solutions of
\be\label{spherevneq}
(\wt\cD_0 v_n)^i = s_n^i,\qq n=1,2,\dots,\qq (\wt\cD_0 v)^i \equiv \g^{jk}\g^{i\l} D_j (P_0v)_{k\l}~,
\ee
where the sources are given as 
\be
s_1^i=- \g^{jk}\g^{i\l}D_j(P_1 v_{n-1})_{k\l},\qq s_n^i=- \g^{jk}\g^{i\l} \le( D_j(P_1 v_{n-1})_{k\l} + D_j(P_2 v_{n-2})_{k\l}\ri),\quad n\geq 2~.
\ee
The operator $-\wt\cD_0$ is Hermitian and non-negative as we have
\be
{-}\int d^dx\sqrt{\g}\,\g_{ij}v^i(\wt\cD_0 v )^j = {1\/2}\int d^d x \sqrt{\g} \,\g^{ik}\g^{j\l} (P_0 v)_{ij} (P_0 v)_{k\l}\geq 0, 
\ee
using integration by parts. This can only vanish when $P_0v =0$ so that $v$ is a CKV. This shows that the only zero modes of $\wt\cD_0$ are the CKVs. A similar argument shows that for any vector field $v$, $\wt\cD_0 v$ is always orthogonal to the CKVs. Note that this was used by York in \cite{doi:10.1063/1.1666338} to prove the existence of his decomposition.

As a result, the operator $-\wt\cD_0$ preserves the space of vector fields orthogonal to the CKVs and is strictly positive on that space. We can see that the sources $s^i_n$ belong to that space. Indeed for any CKV $\z$, we have
\bea
\int d^d x\sqrt{\g}\,\g_{ij} \z^i s^j_n  \= \int d^d x\sqrt{\g}\, \g^{ik}\g^{j\l}D_i \z_j \le( (P_1 v_{n-1})_{k\l}+(P_2 v_{n-2})_{k\l}\ri) \-
\= {1\/2}\int d^d x\sqrt{\g}\, \g^{ik}\g^{j\l} (P_0\z)_{ij} \le( (P_1 v_{n-1})_{k\l}+(P_2 v_{n-2})_{k\l}\ri) \-
\= 0~,
\eea
using integration by parts, tracelessness and symmetry of $(P v)_{ij}$, and the fact that $P_0\z=0$.

As the operator $\wt\cD_0$ is invertible on the space of vector fields orthogonal to the CKVs,  the corrections $v_n$ in \eqref{spherevneq} always exist and are unique. An explicit representation can be written by decomposing the sources in eigenvectors $\{u_k\}$ of $\wt\cD_0$:
 \be
s_n^i = \sum_k c_k u_k^i,
 \ee
 where $\wt\cD_0 u_k =- \la_k u_k$ with $\la_k>0$. This is well-defined because $\wt\cD_0$  is an elliptic operator on a compact manifold and hence has a discrete spectrum. The solution can then be written as
 \be
v_n^i = -\sum_k {c_k\/\la_k} u_k^i~.
 \ee
We can check that the $\r{SO}(1,d+1)$ algebra is satisfied after using the modified Lie bracket \eqref{modifiedbracket} which takes into account the transformation of the metric-dependent corrections.

More generally, we expect that for a large class of gauge-fixing conditions, $\r{SO}(1,d+1)$ should always be the residual symmetry group. This is because the CKVs preserve the background metric and it should be always possible to correct them so that they preserve the gauge conditions.

The advantage of the transverse-traceless gauge used in the main text is that translations and dilatations are realized linearly. This results in simple symmetries for cosmological correlators and simplifies the proof of the holography of information. In a different gauge, the symmetries of cosmological correlators relate correlators of different orders. We expect that the holography of information will still hold in alternative gauges, since given the set of all-order cosmological correlators in a region, the residual symmetries can be used to obtain correlators outside that region.

\section{Orthonormal basis of conformal blocks \label{appconformal}}

In this Appendix, we explain that for free fields in the nongravitational limit, the quantum gravity Hilbert space admits a basis in terms of conformal blocks or conformal partial waves. Moreover we will see that the Higuchi inner product is the natural inner product studied in the CFT literature.

We consider a set of free massive scalar fields $\chi_k$ in the principal series so that they have dimensions
\be
\D_k = {d\/2}+ i \nu_k~,\qq k=1,2\dots~,
\ee
with $\nu_k$ is real. We can define a basis of dS invariant states following section \ref{subsecnormnongrav} as 
\be
|\psi \rn = \int d^dx_1\dots d^d x_n\,\psi(x_1,\dots,x_n) :\chi_1(x_1)\dots\chi_n(x_n): |0\rn~,
\ee
where, as in the main text,  $|0\rn$ is the Hartle-Hawking state. Note that we have redefined the basis by replacing the product of operators by its normal-ordered product which simply corresponds to taking a specific linear combination of the basis elements \eqref{freestates}. 

We must take $\psi(x_1,\dots,x_n)$ to transform appropriately under the conformal symmetry so that $|\psi\rn$ is dS invariant. This corresponds to taking $\psi(x_1,\dots,x_n)$ to have the symmetries of a CFT correlator
\be
\psi(x_1,\dots,x_n) \sim \ln O_1(x_1)\dots O_n(x_n) \rn_\r{CFT}~,
\ee
where $O_k(x)$ is a local operator of dimension $d-\D_k$ in a CFT$_d$. This implies that $\psi$ can be decomposed as a sum of conformal blocks or conformal partial waves. In the example of $n=4$, we have the decomposition
\be
\psi(x_1,\dots,x_4)=\sum_{\D,J} c_{\D,J} \Psi_{\D,J}^{\D_1,\dots,\D_4}(x_1,\dots,x_4)~,
\ee
where the conformal partial waves $\Psi_{\D,J}^{\D_i}$ are linear combinations of conformal blocks.  (See \cite{Simmons-Duffin:2017nub} for details.)

In the principal series, the complex conjugate operator $\chi_k^\ast$ has the conjugate dimension $\D_k^\ast = d-\D_k$ and conformal symmetry implies that we have \cite{Sun:2021thf, Hogervorst:2021uvp}
\be
\ln 0| \chi_k(x)^\ast \chi_k(x')|0\rn_{\text{QFT}} =\d^{(d)}(x-x')~.
\ee
This can be derived for example from the asymptotic limit of de Sitter Green's functions. The Higuchi inner product then takes the form
\be
\ln \psi|\psi\rn = {\text{vol}({\sodminone}) \over \text{vol}(\confgrp)} \int d^d x_1 \dots d^d x_n  \,|\psi(x_1,\dots,x_n)|^2~.
\ee
This is actually  the natural inner product on conformal partial waves. In the example of $n=4$, we have the orthogonality relation 
\be
\ln \Psi_{\D,J}^{\D_i},\Psi_{\bar\D,J}^{\bar\D_i} \rn= {\text{vol}({\sodminone}) \over \text{vol}(\confgrp)} \int {d^dx_1\dots d^d x_4} \Psi_{\D,J}^{\D_i}(x_i)\Psi_{\bar\D',J'}^{\bar\D_i}(x_i) = n_{\D,J} 2\pi \d_{J,J'}\d(\nu-\nu')~,
\ee
where we have written $\D = {d\/2}+i\nu,\bD' = {d\/2}-i\nu'$ with $\nu,\nu'\geq 0$ and the normalization constant $n_{\D,J}$ is the one given in  \cite{Simmons-Duffin:2017nub} multiplied with an additional factor of $\text{vol}(\sodminone)$ to match our convention. This appeared recently in \cite{Caron-Huot:2017vep,Simmons-Duffin:2017nub,Karateev:2018oml,Chen:2019gka} following earlier work \cite{Mack:1974jjo, Dobrev:1977qv}.  The case $n=4$ has been most studied in the CFT literature but we expect that similar results exist for all $n$. This implies that conformal partial waves provide an orthonormal basis for the quantum gravity Hilbert space of free fields in dS$_{d+1}$. 

Semi-classical dS$_3$ gravity can be formulated as a Chern-Simons theory \cite{Witten:1988hc}. So it would be interesting to understand the connection of the construction above to the construction of the Hilbert space of Chern-Simons theory in terms of two-dimensional conformal blocks \cite{Witten:1988hf}.

\section{BRST invariance of inner product}\label{appbrst}
In this section we demonstrate that the correlator
\be
\begin{split}\label{eq:path_integral_prototype}
	(\Psi,A \Psi) &=   \int Dg D\chi\,  D c D \bar{c} \, \delta(g_{ii}-d)\delta (\p_i g_{ij}) |\Psi[g,\chi]|^2 A[g, \chi]  e^{- S_\ghosts}~,
\end{split}
\ee
where $|\Psi[g,\chi]|^2$  and  $A[g,\chi]$ are diffeomorphism and Weyl invariant, enjoys BRST symmetry as is expected of gauge fixed path integrals. In order to show this, we introduce BRST transformation for matter, metric and ghost fields. The BRST operator $\delta_\tB$ that we define below should be distinguished from the BRST operator that would arise if we attempted to implement the gravitational constraints  using the BRST formalism. Rather, it arises when we gauge fix functional integrals like \eqref{normproposal} and \eqref{expecA} in order to define norms and correlators. For this reason, it does not appear that the cohomology of the BRST operator discussed in this section has any particular significance. In this Appendix, for simplicity, we do not consider the fixing of residual gauge.

We will proceed in two steps. First we show BRST invariance of the ghost action containing both diffeomorphism and Weyl ghosts. In the next step, we integrate out the Weyl ghost to obtain the effective ghost action \eqref{ghostaction} and show that the inner product path integral \eqref{eq:path_integral_prototype} with this action is also BRST invariant. (See \cite{Kraemmer:1988yb} for a similar procedure in the context of string theory.)

\subsection{BRST formulation}
We remind the reader that the gauge transformation of the fields under \diffWeyl{} group is given by
\begin{align}
	\delta_{(\xi,\varphi)}\chi &= \delta^\tD_\xi\chi + \delta^\tW_\varphi\chi=\xi\cdot\partial\chi -\Delta\varphi\chi, \\
	\delta_{(\xi,\varphi)}g_{ij} &=\delta^\tD_\xi g_{ij}+ \delta^\tW_\varphi g_{ij} =\nabla_i\xi_j + \nabla_j\xi_i +2\varphi g_{ij}~,
\end{align}
where $\delta^\tD$ and $\delta^\tW$ represent an infinitesimal diffeomorphism and a Weyl transformation respectively. The change in gauge fixing functions under this flow is
\begin{align}
	\delta_{(\xi,\varphi)} (g_{ii}-d) &= 2\nabla_k \xi_k + 2g_{ii}\varphi, \\
	\delta_{(\xi,\varphi)} (\partial_j g_{ij}) &= \partial_j\enc{ \nabla_i\xi_j+\nabla_j\xi_i+2\varphi g_{ij}}~.
\end{align}
From here we can read off the full ghost action as,
\begin{equation}\label{eq:ghost_action_full}
	S_\text{gh}^\text{full} = \int d^dx\, \enc{2g_{ii}\bar{b}b+2\bar{b}\nabla_k c_k + 2\bar{c}^i  \partial_j (g_{ij}b) + \bar{c}^i \partial_j \enc{\nabla_ic_j + \nabla_jc_i}}~,
\end{equation}
where $b,\bar{b},c^i,\bar{c}^i$ are the Weyl and diffeomorphism ghost anti-ghost pairs. \paragraph{Structure constants.} 
Commutators of the gauge group algebra can be given through their action on $\chi$,
\begin{align}\label{eq:diff_weyl_alg_chi}
	[\delta^\tD_\zeta,\delta^\tD_\xi]\chi = \delta^\tD_{[\zeta,\xi]}\chi, \qquad
	[\delta^\tW_\varphi,\delta^\tD_\xi]\chi = -\delta^\tW_{\xi\cdot\partial\varphi}\chi,\qquad
	[\delta^\tW_\varphi,\delta^\tW_\varpi]\chi =0~.
\end{align}
It's easy to check that the same commutation relations hold for the action on $g_{ij}$.
\begin{align}\label{eq:diff_weyl_alg_g}
	[\delta^\tD_\zeta,\delta^\tD_\xi]g_{ij} = \delta^\tD_{[\zeta,\xi]}g_{ij}, \qquad
	[\delta^\tW_\varphi,\delta^\tD_\xi]g_{ij} = -\delta^\tW_{\xi\cdot\partial\varphi}g_{ij},\qquad
	[\delta^\tW_\varphi,\delta^\tW_\varpi]g_{ij} =0~.
\end{align}
Consider the diffeomorphism and Weyl basis $\{\hdelta^\tD_{x^i},\hdelta^\tW_x\}$ defined by
\begin{equation}
	\delta^\tD_\xi = \int d^dx\, \xi^i(x) \hdelta^\tD_{x^i},\qquad \delta^\tW_\varphi = \int d^dx\, \varphi(x) \hdelta^\tW_{x}~.
\end{equation}
Define the structure $f(\_|\_,\_)$ as
\begin{align}
	[\hdelta^\tD_{y^i},\hdelta^\tD_{z^j}] &= \int d^d w \, f(w^k|y^i,z^j) \delta^\tD_{w^k},\\
	[\hdelta^\tW_y, \hdelta^\tD_{z^i}] &= \int d^d w \, f(w|y,z^i) \hdelta^\tW_{w},\\
	[\hdelta^\tW_{z^i},\hdelta^\tW_y] &= \int d^d w \, f(w|z,y) \hdelta^\tW_{w}~.
\end{align}
The notation $f(\_|\_,\_)$ has been overloaded so that its indexed and un-indexed arguments indicate diffeomorphism and Weyl basis indices respectively. From the commutation relations \eqref{eq:diff_weyl_alg_chi} or \eqref{eq:diff_weyl_alg_g} we can read off the structure constants,
\begin{align}
	f(w^k|y^i,z^j) &= \partial_{w^i} \delta(w-z) \delta(w-y) \delta^k_j - \partial_{w^j} \delta(w-y) \delta(w-z) \delta^k_i, \\
	f (w|y,z^i) &= -\delta(w-z) \partial_{w^i} \delta (w-y),\\
	f(w|y,z) &= 0~.
\end{align}
\paragraph{BRST transformation.} Let us rewrite the path integral \eqref{eq:path_integral_prototype} as
\begin{equation}
	(\Psi,A \Psi) = \int Dg D\chi\,  D c D \bar{c}\, DbD\bar{b}\,DB DB^i \, e^{-S_\text{g.i.}-S_\text{g.f.}-S_\text{gh}^\text{full}}, \label{eq:brst_integral}
\end{equation}
where have implemented the gauge fixing delta functions through the Nakanishi-Lautrup fields $B,B^i$ and the gauge fixing action
\begin{align}
	S_\text{g.f} &= i\int d^dx \,\enc{ B (g_{ii}-d) + B^i \partial_j g_{ij}}~,
\end{align}
and indicated the rest of the gauge invariant integrand using $e^{-S_\text{g.i}}$. 

The BRST transformation is
\begin{align}\nt
		\delta_\tB^\theta \chi &= \theta\enc{ c^i\partial_i \chi- b\Delta\chi }, &
		\delta_\tB^\theta g_{ij} &=  \theta \enc{\nabla_ic_j+\nabla_jc_i+2bg_{ij}}, \-
		\delta_\tB^\theta c^i &= \theta c^k\partial_k c^i, &
		\delta_\tB^\theta \bar{c}^i &= - i\theta B^i, \\\label{eq:brst_transformations_full}
		\delta_\tB^\theta b &= \theta c^k\partial_k b, &
		\delta_\tB^\theta \bar{b} &= - i\theta B, \-
		\delta_\tB^\theta B^i &=0,&
		\delta_\tB^\theta B & = 0~.
\end{align}

We have used the following formulae to obtain the ghost field transformations
\begin{align}
	\delta_\tB^\theta c^k(w) &= \frac{\theta}{2} \int d^dy\,d^dz\, f(w^k|y^i,z^j) c^i(y) c^j(z),\\
	\delta_\tB^\theta b^k(w) &= \theta \int d^dy\,d^dz\, f(w|y,z^i) b(y)c^i(z)~.
\end{align}
\paragraph{BRST invariance.} We shall show that the amplitude \eqref{eq:brst_integral} is invariant under the above transformation. Let us define the operation $\delta_\tB$ via $$\delta_\tB^\theta \equiv \theta\delta_\tB.$$ Firstly note that the operator $\delta_\tB$ is nilpotent on \textit{all} variables. That is
\begin{equation}
	\delta_\tB \delta_\tB g_{ij} = \delta_\tB \delta_\tB \chi= \delta_\tB \delta_\tB (\bar{c}^i,c^i,\bar{b},b,B^i,B)=0~.
\end{equation}
This is a group theoretic result which can be easily checked (see for instance \cite{10.1143/PTP.60.272}). This further means
\begin{equation}
	\delta_\tB \delta_\tB \enc{\text{any polynomial in field variables}}=0~.
\end{equation}
The ghost and gauge fixing actions can be rewritten as
\begin{align}
	S_\text{gh}^\text{full} &= \int d^dx \, \enc{\bar{b} \delta_\tB (g_{ii}-d) + \bar{c}^i \delta_\tB (\partial_jg_{ij})},\\
	 S_\text{g.f} &= \int d^dx \, \enc{-\delta_\tB \bar{b} \,(g_{ii}-d) -\delta_\tB \bar{c}^i\, \partial_j g_{ij}}~.
\end{align}
Adding these up,
\begin{equation}
	S_\text{gh}^\text{full} + S_\text{g.f} = -\delta_\tB { \int d^dx \, \enc{\bar{b} (g_{ii}-d) + \bar{c}^i (\partial_jg_{ij})}}~.
\end{equation}
Since the sum is BRST exact, we have
\begin{align}
	\delta_\tB (S_\text{gh}^\text{full} +  S_\text{g.f}) &= 0~.
\end{align}

\subsection{Eliminating the Weyl ghost}\label{app:brst_eliminated}
Now, since the $b,\bar{b}$ ghosts in the action \eqref{eq:ghost_action_full} are non-dynamical, we can simply integrate them out to get the effective ghost action
\begin{align}\begin{split}\label{eq:integrating_weyl}
	e^{-\tilde{S}_\text{gh}}&=\int D b  D\bar{b}\, e^{-S_\text{gh}^\text{full}}\\
	 &= \int D b  D\bar{b}\,e^{ -\int d^dx\encbr{\bar{b}\enc{2g_{ii}b+2\nabla_k c_k} + 2\bar{c}^i  \partial_j (g_{ij}b) + \bar{c}^i \partial_j \enc{\nabla_ic_j + \nabla_jc_i}}} \\
	&= \int D b\,  \delta\enc{-2(g_{ii}b+ \nabla_k c_k)} e^{-\int d^dx\encbr{2\bar{c}^i  \partial_j (g_{ij}b) + \bar{c}^i \partial_j \enc{\nabla_ic_j + \nabla_jc_i}}} \\
	&=  \cfpp \exp \encbr{-\int d^dx\, {\bar{c}^i \partial_j \enc{\nabla_ic_j + \nabla_jc_i - \frac{2}{g_{ii}} g_{ij} \nabla_kc_k}}}.\end{split}
\end{align}
Here $\cfpp = \text{det}(-2 g_{i i})$. As we will see shortly, this is invariant under our new BRST transformation and so $\cfpp$ reduces to an unimportant numerical constant. For these reasons we can drop it from our effective ghost action, and quote
\begin{align}
	S_\text{gh}&= \int d^dx\,\bar{c}^i \partial_j \enc{\nabla_ic_j + \nabla_jc_i - \frac{2}{g_{ii}} g_{ij} \nabla_kc_k}~.
\end{align}
 The gauge fixing part $S_\text{g.f}$ remains the same,
\begin{equation}
	S_\text{g.f} = i\int d^dx \,\enc{ B (g_{ii}-d) + B^i \partial_j g_{ij}}~.
\end{equation}
The new BRST transformation is obtained by replacing $b\to-\frac{1}{g_{ii}}\nabla_kc_k$ in \eqref{eq:brst_transformations_full}.
\begin{align}\nt
		\delta_\tB^\theta \chi &= \theta\enc{c^i\partial_i \chi+ \frac{1}{g_{ii}}\nabla_k c_k\Delta\chi }, &
		\delta_\tB^\theta g_{ij} &=  \theta \enc{\nabla_ic_j+\nabla_jc_i- \frac{2}{g_{\ell\ell}}\nabla_k c_kg_{ij}}, \\\label{eq:brst_transformations_eliminated}
		\delta_\tB^\theta c^i &= \theta c^k\partial_k c^i, &
		\delta_\tB^\theta \bar{c}^i &= -i\theta B^i, &\\\nt
		\delta_\tB^\theta B^i &=0, & 
		\delta_\tB^\theta B &=0.
\end{align}
\paragraph{Nilpotence of $\delta_\tB$.} Since the transformations of $c^i,\bar{c}^i, B, B^i$ are unchanged, their nilpotence is trivially maintained. So we only need to show the nilpotence of the transformations of $\chi$ and $g_{ij}$. 

Firstly we note that
\begin{equation}
	\delta_\tB g_{ii}=0~.
\end{equation}
This gives $\delta_\tB \cN_3 =0$. The modified BRST transformation can thus be interpreted as a diffeomorphism followed by a compensating Weyl transformation which preserves $g_{ii}$. Written explicitly in terms of the ghost field, the transformation is
\begin{equation}
	\delta_\tB g_{ij} = (P_g c)_{ij}~,
\end{equation}
where we define
\begin{equation}
	(P_g c)_{ij} \equiv c^k\partial_kg_{ij}+2 g_{k(i}\partial_{j)}c^k-\frac{2}{g_{mm}}g_{ij}\enc{g_{k\ell}\partial_kc^\ell +\frac{1}{2}c\cdot\partial g_{kk}}~,
\end{equation}
whose gauge-fixed version is \eqref{pdef}.

The $b$-ghost transformation in the full analysis is compatible with the substitution $b\to-\frac{1}{g_{ii}}\nabla_kc_k$ in the new transformation. That is,
\begin{equation}\label{eq:b_compatibility}
	\delta_\tB \enc{-\frac{1}{g_{ii}} \nabla_k c_k} = c^i\partial_i \enc{-\frac{1}{g_{ii}} \nabla_k c_k}~.
\end{equation}
Now for the matter field,
\begin{align}\begin{split}
	\delta_\tB \delta_\tB \chi &= \delta_\tB \enc{c^i\partial_i \chi+ \frac{1}{g_{ii}}\nabla_k c_k\Delta\chi} \\
	&= \delta_\tB c^i \partial_i \chi - c^i\partial_i \delta_\tB\chi + \Delta \delta_\tB \enc{\frac{1}{g_{ii}}\nabla_k c_k} \chi - \frac{\Delta}{g_{ii}} \enc{\nabla_k c_k} \delta_\tB \chi \\
	&= c^j\partial_j c^i \partial_i \chi - c^j\partial_j c^i \partial_i\chi -c^ic^j\partial_i\partial_j \chi - c^j\partial_j \enc{\frac{\Delta}{g_{ii}}\nabla_k c_k}\chi+ \frac{\Delta}{g_{ii}} \nabla_kc_k c^j\partial_j \chi \\
	&\quad  +  \delta_\tB\enc{\frac{\Delta}{g_{ii}}\nabla_k c_k} \chi - \frac{\Delta}{g_{ii}} \nabla_kc_k c_j\partial^j \chi - \frac{\Delta^2}{(g_{ii})^2} \nabla_k c_k\nabla_\ell c_\ell\\
	&= 0.\end{split}
\end{align}
In the third line, the second and last terms cancel out due to antisymmetry of the ghost field and the fourth and sixth terms cancel out due to the relation \eqref{eq:b_compatibility}.
After a slightly more tedious computation of the same for the metric we get
\begin{align}\begin{split}
		\delta_\tB \delta_\tB g_{ij} &= \delta_\tB \enc{c^k\partial_kg_{ij}+ g_{ki}\partial_{j}c^k+ g_{kj}\partial_{i}c^k-\frac{2}{g_{mm}}\nabla_k c_k g_{ij}} \\
		&= -c^k\partial_k \enc{g_{\ell i} \partial_j c^\ell + g_{\ell j}\partial_i c^\ell} +\enc{ c^m\partial_m g_{ki} + g_{\ell i} \partial_k c^\ell} \partial_j c^k \\
		&\quad  + \enc{ c^m\partial_m g_{kj} + g_{\ell j} \partial_k c^\ell} \partial_i c^k + g_{ki} \partial_j \enc{c^\ell\partial_\ell c^k} + g_{kj} \partial_i \enc{c^\ell\partial_\ell c^k} \\
		&= 0.\end{split}
\end{align}
Hence, we can once again make the following assertion for the new $\delta_\tB$: 
\begin{equation}
	\delta_\tB \delta_\tB \enc{\text{any polynomial in field variables}}=0~.
\end{equation}
Also once again,
\begin{align}
	S_\text{gh} &= \int d^dx \, \bar{c}^i \delta_\tB (\partial_jg_{ij}),\\
	S_\text{g.f} &= \int d^dx \enc{ iB(g_{ii}-d) -\delta_\tB \bar{c}^i\, \partial_j g_{ij}}~,
\end{align}
giving
\begin{equation}
	S_\text{gh}+ S_\text{g.f} = -  \delta_\tB \int d^dx \, \enc{\bar{c}^i \partial_j g_{ij}} + i\int d^dx \, B (g_{ii}-d)~.
\end{equation}
The first part is BRST exact, and the other parts depend on $g_{ii}$ and $B$, both of which are BRST closed, thereby yielding
\begin{equation}
	\delta_\tB \enc{S_\text{gh}+ S_\text{g.f}} =0~.
\end{equation}
Since $S_\text{g.i}$ is by definition diffeomorphism and Weyl invariant, this concludes the proof of BRST invariance of the correlator \eqref{eq:path_integral_prototype}.

\section{Phase of the wavefunctional \label{wavephase}}
 In the literature on inflationary cosmology, it is generally assumed that the phase of the wavefunctional, $\Psi[g,\chi]$, is not observable. (See \cite{Maldacena:2015bha} for more discussion.) In accordance with this perspective, in the main text we restricted attention to observables that depended only on the field insertions. Such observables are only sensitive to $|\Psi[g,\chi]|^2$ and not to the phase of the wavefunctional.

On the other hand, from a formal perspective one may wish to study a broader class of observables.  The conjugate momenta, that act on the wavefunctional as $\pi^{i j} = -i {\delta \over \delta g_{i j}}$ and $\pi_{\chi}= -i {\delta \over \delta \chi}$ are sensitive to the phase of the state.  However, the derivative acts on the leading phase-factor in the wavefunctional \eqref{psiderstate} to give a divergent contribution in the large-volume limit. This is the reason these  operators are usually excluded in the study of cosmological correlators.  

To obtain finite quantities we focus on the ``dressed'' momentum operators
\be
\label{dressedmomenta}
\tilde{\pi}^{i j}=e^{i S'[g,\chi]} \pi^{i j} e^{-i S'[g,\chi]}; \qquad \tilde{\pi}_{\chi} = e^{i S'[g,\chi]} \pi_{\chi} e^{-i S'[g,\chi]}
\ee
where $S'[g,\chi] = S[g,\chi] + S_{\cal A}[g,\chi]$ and $S_{\cal A}[g,\chi]$ is any solution to the anomaly equation 
\be \label{saweyl}
i \left(2  g_{ij} {\delta \over \delta g_{ij}} - \Delta \chi {\delta \over \delta \chi} \right) S_{\cal A}[g,\chi] = {\cal A}_d.
\ee
These operators remain finite even in the large volume limit.  We will show that the discussion in the main text easily generalizes to cosmological correlators that include such dressed conjugate momenta.

If $A$ is an operator that involves the dressed conjugate momenta then the expression \eqref{expecA} must be generalized to
\be
\label{generalexpecA}
(\Psi,A \Psi) =  {\cout\/\vdw}\int Dg D\chi\, \Psi^*[g,\chi] A \Psi[g, \chi],
\ee
i.e. the functional derivatives in $A$ act on the wavefunctional on the right but not on its conjugate.

Similarly, one may define a generalization of cosmological correlators. Consider the string of operators
\begin{equation}
\label{generalizedstring}
	\cprod^{p,q,r,s}_{\vI\vJ\vK\vL}(\vec{x}) =h_{i_1j_1}(y_1)\ldots h_{i_pj_p}(y_p)\chi(z_1)\ldots\chi(z_q)\tilde{\pi}^{k_1\ell_1}(u_1)\ldots\tilde{\pi}^{k_r\ell_r}(u_r) \tilde{\pi}_\chi(v_1)\ldots\tilde{\pi}_\chi(v_s).
\end{equation}
Then we define a generalized cosmological correlator
\be
\label{generalizedcc}
\llangle \Psi | \cprod^{p,q,r,s}_{\vI\vJ\vK\vL}(\vec{x}) | \Psi \rrangle_{\text{CC}} 
\equiv  \cout \cfp \int D g D \chi D \bar{c} D c' e^{-S_\ghosts} \delta(g_{ii} - d) \delta(\partial_i g_{i j}) \Psi^*[g, \chi]  \cprod^{p,q,r,s}_{\vI\vJ\vK\vL}(\vec{x}) \Psi[g,\chi]
\ee
We note the following.
\begin{enumerate}
\item
The operators \eqref{dressedmomenta} involve a choice of the functional $S_{\cal A}[g,\chi]$. This choice can be made according to convenience and our discussion is valid for any choice. Given correlators of operators with one choice of $S_{\cal A}$, it is evident that one may obtain correlators of operators corresponding to another choice.
\item
In \eqref{generalizedcc}, all the momenta appear on the right. Since the momenta and the fields satisfy canonical commutation relations, it is clear that knowledge of all such correlators suffices to determine cosmological correlators of operators with other orderings.
\end{enumerate}

By combining the anomaly equation \eqref{saweyl} with \eqref{weylinsummary}, we see that the wavefunctional
\be
\widetilde{\Psi}[g,\chi] = e^{-i S'} \Psi[g,\chi]
\ee
is \diffWeyl{} invariant. A slight manipulation shows that the cosmological correlator can be written as
\be
\begin{split}
&\llangle \Psi | \cprod^{p,q,r,s}_{\vI\vJ\vK\vL}(\vec{x}) | \Psi \rrangle_{\text{CC}} = \cout \cfp \int D g D \chi D \bar{c} D c' e^{-S_\ghosts} \delta(g_{ii} - d) \delta(\partial_i g_{i j}) \\ &\times \widetilde{\Psi}[g,\chi]^* h_{i_1j_1}(y_1)\ldots h_{i_pj_p}(y_p)\chi(z_1)\ldots\chi(z_q) {\pi}^{k_1\ell_1}(u_1)\ldots {\pi}^{k_r\ell_r}(u_r) {\pi}_\chi(v_1)\ldots{\pi}_\chi(v_s) \widetilde{\Psi}[g,\chi]
\end{split}
\ee
Here, note that all the dressed momentum operators have been replaced by their ordinary counterparts and it is the wavefunctional that is dressed instead.

Using the \diffWeyl{} invariance of $\widetilde{\Psi}[g,\chi]$, and by repeating the arguments in subsection \ref{subsecsymm}, we see that cosmological correlators involving the momenta have the following symmetries.
\begin{align} \label{eq:gen_cosmo_scaling}
	\CC{\Psi}{\cprod^{p,q,r,s}_{\vI\vJ\vK\vL}(\lambda\vec{x}+\zeta)} &= \lambda^{-q\Delta-rd-s\bar{\Delta}} \CC{\Psi}{\cprod^{p,q,r,s}_{\vI\vJ\vK\vL}(\vec{x})},\\
	\llangle \Psi | \cprod^{p,q,r,s}_{\vI\vJ\vK\vL}(R \cdot \vec{x}) | \Psi \rrangle_{\text{CC}} & = R^{i'_1}_{i_1} R^{j'_1}_{j_1} \ldots R^{i'_p}_{i_p}R^{j'_p}_{j_p} \times R_{k'_1}^{k_1} R_{\ell'_1}^{\ell_1} \ldots R_{k'_r}^{k_r}R_{\ell'_r}^{\ell_r}  \llangle \Psi | \cprod^{p,q,r,s}_{\vI'\vJ'\vK'\vL'}(\vec{x}) | \Psi \rrangle_{\text{CC}},\label{eq:gen_cosmo_rot}
\end{align}
with $\bar{\Delta} = d - \Delta$. 
This is a straightforward generalization of (\ref{csymshiftscal},\ref{csymrot}) with the corresponding momentum insertions acquiring dual scaling dimensions and the gravity momenta rotating the inverse way.

The result of section \ref{secholinfo} can now be generalized to
\begin{result}
The set of all generalized cosmological correlators of the form \eqref{generalizedcc} in any open region $\reg$ in a state $\Psi$ is sufficient to determine the wavefunctional $\Psi[g,\chi]$. 
\end{result}
We emphasize that the generalized correlators fix not just the magnitude but also the phase of the wavefunctional. 

\bibliography{references}
\end{document}